\documentclass[showpacs,onecolumn,aps,10pt ]{revtex4}
\usepackage{amsfonts}
\usepackage{graphicx}
\usepackage{amsmath}
\usepackage{amssymb}
\usepackage{amsthm}
\usepackage{color}

\def\be{\begin{equation}}
\def\ee{\end{equation}}
\def\bea{\begin{eqnarray}}
\def\eea{\end{eqnarray}}

\newcommand{\f}[2]{\frac{#1}{#2}}

\begin{document}

\title{Vector dark energy models with quadratic terms in the Maxwell tensor
derivatives}
\author{Zahra Haghani$^1$}
\email{z.haghani@du.ac.ir}
\author{Tiberiu Harko$^{2,3}$}
\email{t.harko@ucl.ac.uk}
\author{Hamid Reza Sepangi$^4$}
\email{hr-sepangi@sbu.ac.ir}
\author{Shahab Shahidi$^1$}
\email{s.shahidi@du.ac.ir}
\affiliation{$^1$ School of Physics, Damghan University, Damghan, Iran,}
\affiliation{$^2$ Department of Physics, Babes-Bolyai University, Kogalniceanu Street,
Cluj-Napoca 400084, Romania,}
\affiliation{$^3$Department of Mathematics, University College London, Gower Street,
London WC1E 6BT, United Kingdom,}
\affiliation{$^4$Department of Physics, Shahid Beheshti University, G. C., Evin, Tehran
19839, Iran}

\begin{abstract}
We consider a vector-tensor gravitational model with terms quadratic in the Maxwell tensor derivatives, called the Bopp-Podolsky term. The gravitational field equations of the model and the equations describing the evolution of the vector field are obtained and their Newtonian limit is investigated.  The cosmological implications of a Bopp-Podolsky type dark energy term are investigated for a Bianchi type I homogeneous and anisotropic geometry for two models, corresponding to the absence and presence of the self-interacting potential of the field, respectively.  The time evolutions of the Hubble function, of the matter energy density,  of the shear scalar, of the mean anisotropy parameter, and of the deceleration parameter, respectively, as well as  the field potentials are obtained for both cases by numerically integrating the cosmological evolution equations. In the presence of the vector type dark energy with quadratic terms in the Maxwell tensor derivatives, depending on the numerical values of the model parameters,  the Bianchi type I Universe experiences  a complex dynamical evolution, with the dust Universes ending in an isotropic  phase. The presence of the self-interacting potential of the vector field significantly shortens the time interval necessary for the full isotropization of the Universe.
\end{abstract}

\maketitle

\section{Introduction}
Recent cosmological observations, based initially on the study of the
distant Type Ia Supernovae, have shown that the cosmological paradigm
according to which the Universe must decelerate due to its own gravitational
attraction is not correct and that the Universe has experienced a transition to a late
time, de Sitter type accelerated phase \cite{1n,2n,3n,4n}. These
observations have triggered a deep revision of our
understanding of the cosmological dynamics and of its theoretical basis,
general relativity. To explain the current observations in cosmology many
theoretical ideas and suggestions have been put forward to address the
intriguing facts revealed by the complex observational study of the
Universe. From both theoretical and observational points of view there is a
general consensus, which may be referred to as ``the standard explanation of the late
time acceleration,'' according to which observations can be easily explained
once we assume the existence of a mysterious and dominant component in the
Universe, called \textit{dark energy}, and  is fully responsible for
the observed dynamics in the late phases of the evolution of the Universe.

Another important cosmological result, based on the combination of data
from the observations of high redshift supernovae, the WMAP satellite,
and the recently released Planck data, convincingly show that the
location of the first acoustic peak in the power spectrum of the CMBR
(Cosmic Microwave Background Radiation) is entirely consistent with the
important prediction of the inflationary model for the total density
parameter $\Omega $ of the Universe, according to which at the end of
inflation $\Omega =1 $. The important cosmological parameter $w=p/\rho $,
where $p$ is the total pressure and $\rho $ is the total density of the
Universe is also strongly constrained by cosmological observations
which provide detailed evidence for the behavior of the equation of state
of the cosmological fluid, constraining the parameter $w$ as lying in the
range $-1\leq w <-1/3$ \cite{acc}.

These large number of cosmological observations have led to the formulation
of the $\Lambda $CDM paradigm according to which, in order to explain the
cosmological evolution, one assumes that the Universe is filled with two main
components representing around 95\% of its content; cold (pressureless)
dark matter (CDM) and dark energy (DE), having a negative pressure.
The contribution of the CDM component to the total density
parameter of the Universe is of the order of $\Omega _{m}\sim 0.3$ \cite{P2}.
From a theoretical point of view the necessity to consider dark
matter is mainly required by the necessity of explaining the unusual
behavior of the galactic rotation curves as well as the formation of the
large scale structure. On the other hand, dark energy is considered as
representing the major component of the ``chemical'' composition of the
Universe, giving a contribution to the total density parameter of the order
of $\Omega _{DE}\sim 0.7$. Dark energy is the major cause determining the
recent, de Sitter type, acceleration of the Universe, as confirmed by the
study of high redshift type Ia supernovae \cite{acc}. The search for an
explanation of the physical (or geometrical) nature and  properties
of dark energy has opened  a very active field of research in
cosmology and theoretical physics which in turn has led to a myriad
of different DE models, for reviews of DE models see, for
example, \cite{PeRa03,new1,Pa03,DEreviews, Od,LiM, Mort,Amend}.

One interesting possibility for explaining DE which has been
intensively investigated is based on a number of cosmological models in which the ``chemical''
composition of the Universe consists of a mixture of two major components;
cold dark matter and a slowly-varying, spatially inhomogeneous component,
called the quintessence \cite{8n}. In this
scenario the baryonic matter plays a negligible role with a minimal
influence on late time cosmological dynamics. From a formal theoretical
point of view and based on some particle physics results, quintessence type
cosmological models can be implemented by assuming that dark energy is the
energy associated with a scalar field $Q$ with a self-interacting potential
$V(Q)$ \cite{Fa04}. During the cosmological evolution, when the potential
energy density $V(Q)$ of the quintessence field becomes greater than the
kinetic energy density,  the thermodynamic pressure $p=\dot{Q}%
^{2}/2-V(Q) $, associated with the quintessence $Q$-field, becomes negative.
The cosmological and astrophysical properties of the quintessential
cosmological models have been intensively investigated in the literature,
for a recent review see \cite{Tsu}. Quintessence models differ from
cosmological models of the standard general relativity including the cosmological
constant since they imply that the equation of state of the quintessence
field varies dynamically with cosmic time \cite{11n}. A number of
alternative cosmological models, called $k-$essence, where the
late-time acceleration of the Universe is driven by the kinetic energy of
the scalar field have also been proposed \cite{kessence0}.

Another possibility to explain the recent acceleration of the Universe and
the nature of dark energy is provided by scalar fields $\phi $ that are
minimally coupled to gravity via a negative kinetic energy. An interesting
property of these fields is that they allow for values of the equation of state parameter, $w$,
of dark energy to vary in such a way as to have $w<-1$. These types of scalar
fields are known as phantom fields,  proposed as an
explanation for the late time acceleration of the Universe in \cite{phan1}.
The energy density and the pressure of  phantom scalar fields are given
by $\rho _{\phi}=-\dot{\phi}^2/2+V\left(\phi \right)$ and $p _{\phi}=-\dot{%
	\phi}^2/2-V\left(\phi \right)$, respectively. Phantom cosmological models
for dark energy have been investigated in detail in \cite{phan2,phan3, phan4,new2}.
Some recent cosmological observations seem to support the interesting
result that at some instant during the evolution of the Universe the value of
$w$ representing  dark energy equation of state may have crossed the
standard value $w= -1$, hence entering a de Sitter type expansion
with a cosmological constant $\Lambda $. This intriguing cosmological phenomenon is called \textit{the
	phantom divide line crossing} \cite{phan4}. The crossing of the phantom
divide line was investigated in the case of scalar field models with cusped
potentials in \cite{phan3}. The phantom divide line crossing can also be
explained in cosmological models where dark energy is represented by a
scalar field, which is non-minimally coupled to gravity \cite{phan3}.

A different line of research on dark energy is based on the assumption that
instead of interpreting dark energy as a specific physical field, the
cosmological dynamics of the Universe can be understood as a modification of
the gravitational force itself. By following this line of thought one can
assume that at very large cosmological scales general relativity cannot
describe the dynamical evolution of the Universe, and therefore the
acceleration of the Universe is related to an intrinsic change of the
gravitational interaction. A plethora of modified gravity models, based on
different extensions of general relativity, like, for example,  $f(R)$ gravity (in
which the gravitational action is an arbitrary function of the Ricci scalar $%
R$) \cite{Bu70} and mimetic-$f(R)$ gravity models \cite{mime}, the $f\left( R,L_{m}\right) $ model (where $L_{m}$ is the
matter Lagrangian) \cite{Har1}, $f(R,T)$ modified gravity models (where $T$
denotes the trace of the energy-momentum tensor) \cite{Har2}, the
Weyl-Cartan-Weitzenb\"{o}ck (WCW) model \cite{WCW}, hybrid metric-Palatini $%
f(R,\mathcal{R})$ gravity models (where $\mathcal{R}$ is the Ricci scalar
formed from a connection independent of the metric) \cite{Har3}, $%
f\left(R,T,R_{\mu \nu }T^{\mu \nu }\right)$ type models, where $R_{\mu \nu
} $ is the Ricci tensor and $T_{\mu \nu }$ is the matter energy-momentum
tensor, respectively \cite{Har4}, the Eddington-inspired Born-Infeld theory
\cite{EIBI}, $f(\tilde{T},\mathcal{T})$ gravity \cite{HT}, implying
coupling between  torsion scalar $\tilde{T}$ and  trace of the matter
energy-momentum tensor, or vector Gauss-Bonnet theory \cite{vgb}, have been recently proposed in the literature. The
cosmological and astrophysical properties of these models have been
extensively investigated. For a recent review of the generalized
gravitational models with non-minimal curvature-matter coupling $f\left(
R,L_{m}\right)$ and $f(R,T)$ type see \cite{Revn}. For a review of hybrid
metric-Palatini gravity see \cite{Revn1}. Modified gravity models can
provide convincing theoretical explanations for the late time acceleration of the
Universe without advocating the existence of dark energy and can also
offer some alternative explanations for the nature of dark matter.

From a field theoretical point of view however, despite the great
success of the scalar field dark energy models, the possibility that dark
energy has a more complex structure than allowed by the simple scalar field
model cannot be ignored \textit{a priori}. One promising direction in the
analysis of dark energy is represented by models in which dark energy is
described by a vector or Yang-Mills type field which may also couple,
minimally or non-minimally, to gravity. The simplest action for a Yang-Mills
type dark energy model is \cite{v1,new5}
\begin{align}
S_{V} =\int d^{4}x\sqrt{-g}\Big(\frac{R}{2}-\frac{1%
}{4}F_{\mu \nu }^{a}F^{a}{}^{\mu \nu }+V(A^{2})+L_{m}\Big),  \label{actv}
\end{align}%
where $A_{\mu }^a$, $a=1,2,...,n$ are the potentials of the Yang-Mills
field, $F_{\mu \nu }^{a}=\nabla _{\mu }A_{\nu }^{a}-\nabla _{\nu }A_{\mu
}^{a} $, $\nabla _{\mu }$ is the covariant derivative with respect to the
metric, $A^{2}$ is defined as $A^{2}=g^{\mu \nu }A_{\mu }^{a}A_{\nu
}^{a}$ and $V(A{}^{2})$ represents a self-interacting potential,
explicitly violating gauge invariance. In the action given by Eq.~(\ref{actv}%
) there are three vector fields describing dark energy. Hence Eq.~(\ref{actv}%
) generalizes the Einstein-Maxwell type single vector field dark energy
model. The astrophysical and cosmological applications of the single or
Yang-Mills type vector dark energy models have been comprehensively investigated
in \cite{v2}.

Extended vector field dark energy models where the vector field is
non-minimally coupled to the gravitational field can also be constructed
\cite{v3}. The action for such a non-minimally coupled vector dark energy
model is given by
\begin{align}\label{1}
S =\int d^{4}x\sqrt{-g}\Bigg[&\frac{R}{2}-\frac{1}{4}F_{\mu \nu
}F^{\mu \nu }-\frac{1}{2}\mu _{\Lambda }^{2}A_{\mu }A^{\mu }+\omega A_{\mu }A^{\mu }R+\eta A^{\mu }A^{\nu }R_{\mu \nu }+L_{m}\Bigg],
\end{align}%
where $A^{\mu }\left( x^{\nu }\right) $, $\mu ,\nu =0,1,2,3$ is the
four-potential of the vector type dark energy, which couples non-minimally
to gravity, and $\mu _{\Lambda }$ is the mass of the massive cosmological
vector field, respectively. The constants $\omega $ and $\eta $ are
dimensionless coupling parameters, while the vector dark energy field tensor
is defined as $F_{\mu \nu }=\nabla _{\mu }A_{\nu }-\nabla _{\nu }A_{\mu }$.

Inspired by the possible analogy between dark energy and some condensed
matter concepts, a so called superconducting type dark energy model was
proposed in \cite{SupracondDE}. This model describes the spontaneous
breaking of  U(1) symmetry of the ``electromagnetic'' type dark energy, and
is described by the action
\begin{eqnarray}  \label{s1}
S=\int d^{4}x\sqrt{-g}\Bigg[ \frac{R}{2}-\frac{1}{4}F_{\mu \nu }F^{\mu \nu }
-\frac{%
	\lambda}{2} \left( A^{\nu }-\nabla ^{\nu }\phi \right) \left( A_{\nu }-\nabla _{\nu
}\phi \right) +V\left( A^2,\phi \right) -\frac{\alpha}{2} j^{\nu }\left( A_{\nu }-\nabla _{\nu }\phi
\right) +L_{m}\left( g_{\mu \nu },\psi \right) \Bigg],
\end{eqnarray}%
where $\lambda $ and $\alpha $ are constants, $L_{m}\left( g_{\mu \nu },\psi
\right) $ is the Lagrangian of the total (ordinary baryonic plus dark)
matter, and $j^{\mu }=\rho u^{\mu }$ is the total mass current, where $\rho $
is the total matter density (including dark matter), and $u^{\mu }$ is the
matter four-velocity. This model can also be interpreted and understood as
unifying, in a single formalism, the scalar and vector dark energy models. The
predictions of the superconducting dark energy model have been compared
with observations in \cite{supobs}.

It is the goal of the present paper to consider a vector-tensor type model
of dark energy, based on the analogy with Bopp-Podolsky electrodynamics. The
Bopp-Podolsky theory was first suggested by Bopp \cite{Bopp}, and  was
independently reobtained by Podolsky \cite{Podolsky}. The Bopp-Podolsky
theory retains linearity of the field equations but introduces
higher-derivative terms proportional to the parameter $m^2$ where $m$,
having the physical dimensions of mass, is a new hypothetical
fundamental constant of Nature. For $m\rightarrow\infty$ the Maxwell-Lorentz
theory, and the Maxwell equations, are retained. The Bopp-Podolsky theory
is formulated in terms of an action functional from which the field
equations, which are of fourth order in the electromagnetic potential, are
derived. However, as noted by both Bopp and Podolsky, in a certain gauge
these fourth-order equations are equivalent to a pair of second-order
equations \cite{Bopp,Podolsky}. Different aspects of the Bopp-Podolsky type
extension of classical electrodynamics were investigated in \cite{applBopp}.

To this and other ends, we start from the analogy with the Bopp-Podolsky electrodynamics and introduce a vector-tensor gravitational model where the action for the minimally coupled vector field also contains additional terms, quadratic in the Maxwell tensor derivatives. These terms correspond to the covariant form of the action of the Bopp-Podolsky electrodynamics. Moreover, a term describing the non-minimal coupling between the cosmological mass current and the four-potential of the vector field is also added to the action. The possible existence of a self-interaction potential of the vector field is also considered. From a cosmological point of view we propose to interpret the vector field as describing the dark energy component of the Universe,  which is responsible for the late, de Sitter type acceleration of the Universe. We obtain the gravitational field equations of this vector dark energy model as well as the equations describing the evolution of the vector field. We investigate the Newtonian limit of the model and  show that the Poisson equation as well as the Bopp-Podolsky electrodynamics can be recovered for weak fields.

The cosmological implications of this vector type dark energy model are investigated for a Bianchi type I homogeneous and anisotropic geometry. Two cases are investigated in detail, the evolution of the Universe with and without the self-interacting potential of the field, respectively.  In both cases the evolution of the Hubble function, of the matter energy density, of the shear scalar, of the anisotropy parameter, of the deceleration parameter, and of the field potential are analyzed in detail. To escribe the matter content of the Universe we adopt the radiation fluid and the dust matter equations of state. We find that in the presence of the vector type dark energy with quadratic terms in Maxwell tensor derivatives the anisotropic Universe experiences a complex dynamical behavior, with the dust Universes ending in an isotropic  stage, a result which is independent on the presence or absence of the self-interaction potential of the field.

The present paper is organized as follows. The field equations of the Bopp-Podolsky type vector-tensor gravitational model are derived in Section~\ref{sect2} and their Newtonian limit is also investigated. The cosmological implications of the model are investigated in Section~\ref{sect3}, where the cosmological dynamics of a Bianchi type I geometry is analyzed for both models with and without self-interaction potential of the vector field, and for two different equations of state of the cosmic matter. We discuss and conclude our results in Section~\ref{sect4}.

\section{Bopp-Podolsky type vector dark energy models}\label{sect2}

We first start by briefly introducing the basic theoretical
ideas of the Bopp-Podolsky type electrodynamics in its standard formulation
in Minkowski geometry. Then, by adopting, as a starting point, the view
that higher order derivatives of the Maxwell tensor may play a significant
role in vector type models of dark energy, we introduce the gravitational
action for such a theoretical model. The gravitational field equations as
well as the equations of the vector field are derived from the action together
with an equation representing the covariant conservation of the energy-momentum tensor.

\subsection{The Bopp-Podolsky model of electrodynamics}

The Lagrangian density from which Maxwell's equations can be obtained by the
usual variational principle is \cite{LaLi}
\begin{equation}
L_{0}=-\frac{1}{4}F_{\mu \nu }^{2}+j_{\mu }A^{\mu },  \label{L1}
\end{equation}%
where we use a system of units with $c=1$. In the above equation $F_{\mu
	\nu} $ is the Maxwell electromagnetic field tensor, $A_{\mu}$ is the
four-vector potential of the field while $j_{\mu}$ denotes the
electromagnetic current. This Lagrangian is a function of the field
variables and of their first derivatives. There is no reason why we should
restrict ourselves to only first derivatives in the action and it therefore
seems natural to try a generalization of Eq. (\ref{L1}) of the form \cite%
{Bopp,Podolsky}
\begin{equation}
L=L_{0}\left( F_{\mu \nu },\frac{\partial F_{\mu \nu }}{\partial x^{\rho }}%
\right) +j_{\mu }A^{\mu }.  \label{L2}
\end{equation}

The usual variational principle applied to this Lagrangian leads to the
field equations
\begin{equation}
\partial^\nu f_{\mu \nu }=j_{\mu },  \label{f1}
\end{equation}%
where
\begin{equation}
f^{\mu \nu }=-2\left[ \frac{\partial L_{0}}{\partial F_{\mu \nu }}-\frac{%
	\partial }{\partial x^{\rho }}\left( \frac{\partial L_{0}}{\partial \left(
	\partial F_{\mu \nu }/\partial x^{\rho }\right) }\right) \right] .
\label{f2}
\end{equation}

The simplest choice for $L_{0}$, as proposed by Bopp and Podolsky \cite%
{Bopp,Podolsky}, is
\begin{equation}
L_{0}=-\frac{1}{4}\left[ F_{\mu \nu }^{2}-\frac{1}{m^2}\left(
\frac{\partial F_{\mu \nu }}{\partial x^{\rho }}\right) ^{2}\right] ,
\label{L3}
\end{equation}
where $m$ is a new fundamental constant with mass dimension $1$.
Using Eq.~(\ref{L3}) in Eq.~(\ref{f2}) we obtain
\begin{equation}
f_{\mu \nu }=F_{\mu \nu }+\frac{1}{m^{2}}\square F_{\mu \nu },
\end{equation}
so that the field Eq. (\ref{f1}) becomes
\begin{equation}
\left( \square +m^{2}\right) U_{\mu }=- j_{\mu },
\end{equation}
where
\begin{equation}
U_{\mu }=-\frac{1}{m^{2}}\partial^\nu F_{\mu \nu }.  \label{f3}
\end{equation}
$U_{\mu }$ has the property
$
\partial^\mu U_{\mu }=0.
$
By defining
\begin{equation}
U_{\mu \nu }=f_{\mu \nu }-F_{\mu \nu }=\frac{1}{m^{2}}\square F_{\mu \nu
},
\end{equation}
we obtain
\begin{equation}
\partial^\nu U_{\mu \nu }=j_{\mu }+m^{2}U_{\mu }.
\end{equation}
If we impose the condition
$
\partial_\mu A^\mu=0,
$
Eq. (\ref{f3}) becomes
\begin{equation}
\square A_{\mu }-m^{2}U_{\mu }=0.
\end{equation}
By introducing the four-potential
$
a_{\mu }=A_{\mu }+U_{\mu },
$
we obtain
\begin{align}
f_{\mu \nu }=\frac{\partial a_{\nu }}{\partial x^{\mu }}-\frac{\partial
	a_{\mu }}{\partial x^{\nu }}.
\end{align}%

To summarize, an interesting result in the Bopp-Podolsky theory is that the
electromagnetic field equations, the potentials and the fie1d strengths can
be written as the difference, respectively, of the potentials and field
strengths of two distinct fields
\begin{equation}
A_{\mu }=a_{\mu }-U_{\mu },
\end{equation}
\begin{equation}
F_{\mu \nu }=f_{\mu \nu }-U_{\mu \nu }.
\end{equation}
These two fields are described by two sets of separate field equations, with
the first set corresponding to the standard Maxwell equations, while the
second set represents Proca type field equations for particles with mass $%
m$. This result also provides a physical interpretation of the new
fundamental constant $m$. However, the mass term appears with a wrong sing in the equation of motion,
signaling that the massive vector field is a ghost. In order to make it clear, let us rewrite the Lagrangian
\eqref{L3} in the Lorentz gauge with the result
\begin{align}
L_0=\f12 A_\mu (\Box+\f{\Box^2}{m^2})A^\mu.
\end{align}
One can easily check that the above Lagrangian is identical to
\begin{align}
L_{alt}=\f12 A_\mu\Box A^\mu-A_\mu\Box B^\mu-\f{1}{2}m^2B^\mu B_\mu.
\end{align}
Now using the transformation $A_\mu\rightarrow A_\mu+B_\mu$, one  obtains
\begin{align}\label{zah1}
L_{alt}\rightarrow\f12 A_\mu\Box A^\mu-\f12 B_\mu\Box B^\mu-\f{1}{2}m^2 B^\mu B_\mu.
\end{align}
It is now seen that the kinetic term of the massive vector field appears with a wrong sign, signaling that the massive vector field is a ghost. In order to make the theory healthy at the background level, one should make the massive ghost non-dynamical. In this paper, we will consider the cosmology of this model, so the Maximum energy scale of our theory is $H_0$. By assuming that the ghost mass is larger than the energy scale of our theory, the ghost mass does not have any dynamics at length scales smaller than $H_0^{-1}$ which is the desired range of doing cosmology. This means that the Kinetic energy of the ghost field is much less than its potential energy. Noting that the ghost mass squared is equal to $m^2$, one can see that for values $m\gg H_0$, the ghost becomes non-dynamical. This is what we consider in what follows.
\subsection{Bopp-Podolsky type vector dark energy models}

In the following we assume that the vector dark energy can be described by a
Bopp-Podolsky type model, with the action given by
\begin{align}
S = \int \sqrt{-g} d^4x\bigg( \frac{1}{16\pi}R-\frac{1}{4}%
F^{\mu\nu}F_{\mu\nu}+V(A^2)+\frac{1}{4m^2}\nabla ^\rho F^{\mu\nu} \nabla _\rho F_{\mu\nu} - \beta A
_\mu j^\mu \bigg)+S_m,
\end{align}
where $A_\mu$ is the dark energy potential related to dark energy field
strength by
\begin{equation}  \label{eq:FA}
F_{\mu\nu} \, = \, \nabla _\mu A_\nu \, - \, \nabla _\nu A_\mu,
\end{equation}
where $A^2=A^\mu A_\mu$, $j^\mu=\rho u^\mu$ is the cosmological matter 4-vector
current, $m$ is a constant with dimension of mass and $S_m$ is the action for ordinary matter. We have also added to
the gravitational action the self-interacting potential $V\left(A^2\right)$
of the vector field, and we have allowed for the possibility of a direct
coupling between the matter current $j_{\mu}$ and  dark energy vector
potential $A^{\mu}$, with the strength of the coupling described by the
constant $\beta $.

Varying the action \eqref{eq:FA} with respect to  dark energy potential $%
A_\mu$ and the metric $g_{\mu\nu}$ we have
\begin{align}  \label{vec}
\nabla_\mu F^{\mu\nu}+\f{1}{m^2}\nabla_\mu\Box F^{\mu\nu} +2V^\prime(A^2)
A^{\nu}=\beta J^\nu,
\end{align}
where a prime indicates  derivative with respect to the argument and
\begin{widetext}
	\begin{align}\label{met}
	G_{\mu\nu}&-8\pi\left(F_{\mu\alpha}F_\nu^{~~\alpha}-\f14 F_{\alpha\beta}F^{\alpha\beta}g_{\mu\nu}\right)+\f{4\pi}{m^2}\bigg[\nabla_\mu F^{\alpha\beta}\nabla_\nu F_{\alpha\beta}-\f12\nabla_\rho F_{\alpha\beta}\nabla^\rho F^{\alpha\beta} g_{\mu\nu}+2\nabla_{\alpha}F_{\beta(\nu}\nabla^\alpha F^{\beta}_{~~\mu)}
	\nonumber\\&+2\nabla_\alpha \big(F^{\alpha\beta}\nabla_{(\mu} F_{\nu)\beta}+F_{\beta(\mu}\nabla_{\nu)}F^{\alpha\beta}+F_{\beta(\mu}\nabla^\alpha F_{\nu)}^{~~\beta}\big)\bigg]+16\pi \left[A_\mu A_\nu V^\prime(A^2) -\f{1}{2}V(A^2) g_{\mu\nu}\right]\nonumber\\&~~~~~~~~-8\pi\beta A_\alpha u^\alpha p(u_\mu u_\nu+g_{\mu\nu})=8\pi T_{\mu\nu}.
	\end{align}
\end{widetext}
At this point, a note about the variation of the term $A_\alpha j^\alpha\equiv\rho u^\alpha
A_\alpha$ is in order. The variation of the energy momentum tensor can be
written as (see Appendix \ref{app})
\begin{equation*}
\delta\rho=\frac{1}{2}(p+\rho)(u_\mu u_\nu +g_{\mu\nu})\delta g^{\mu\nu},
\end{equation*}
while the variation of the four-velocity of the particle is
\begin{equation*}
\delta u^\alpha=-\frac{1}{2 } u^\alpha u_\mu u_\nu \delta g^{\mu\nu}.
\end{equation*}
Putting all these results in the variation of $A_\mu j^\mu$, one can see
that $\rho$ dependence vanishes from the metric equation of motion. In
the following we will assume that the energy momentum tensor of  ordinary
matter is that of a perfect fluid
\begin{align}
T_{\mu\nu}=(\rho+p)u_\mu u_\nu+pg_{\mu\nu}.
\end{align}
In order to write the equation of the vector field in a form similar to the one in  Bopp-Podolsky electrodynamics, we introduce a new auxiliary vector field $U_{\mu}$, defined
as
\begin{equation*}
U_\nu=-\frac{1}{m^2}\nabla^\mu F_{\mu\nu}.
\end{equation*}
Then the vector field equation \eqref{vec} reduces to two coupled
differential equations for $A_\mu$ and $U_\mu$ as
\begin{align}
\Box A^\nu-R^{\alpha\nu}A_\alpha=\nabla^\nu(\nabla_\alpha A^\alpha)-m^2U^\nu,
\end{align}
and
\begin{align}
(\Box+m^2)U^\nu&-\f{1}{m^2}\nabla^\rho(R_{\alpha\rho}F^{\alpha\nu})-\f{1}{m^2}\nabla_\rho
R^{\rho\mu\alpha\nu}F_{\mu\alpha}-\f{2}{m^2}R^{\rho\mu\alpha\nu}\nabla_\rho F_{\mu\alpha}-2 V^\prime(A^2) A^{\nu}=-\beta
J^\nu,
\end{align}
respectively. The conservation of the energy momentum tensor is now obtained by
taking the covariant divergence of the metric field equation. After some
algebra, one  finds
\begin{align}  \label{conE}
\nabla^\mu T_{\mu\nu}=&\beta F_{\mu\nu}J^\mu +2 A_\nu \nabla^\mu(V^\prime
A_\mu) -\beta h_{\mu\nu}\nabla^\mu\big((A_\alpha u^\alpha)p\big)-\beta(A_\alpha u^\alpha)p(\theta u_\nu+a_\nu),
\end{align}
where $\theta=\nabla_\mu u^\mu$ is the expansion parameter, $%
a^\mu=u^\nu\nabla_\nu u^\mu$ is the acceleration and $h_{\mu\nu}=g_{\mu%
	\nu}+u_\mu u_\nu$.

By taking the covariant derivative of equation \eqref{vec} we obtain
\begin{equation}\label{sh1}
\nabla_\mu (A^\mu V^\prime(A^2))=\frac{\beta}{2} \nabla_\mu J^\mu.
\end{equation}

\subsection{The Newtonian limit}

In the following we consider the weak field limit of the Bopp-Podolsky type
vector dark energy model for a static source, i.e., the Newtonian limit. In
this case the only non-zero component of the energy-momentum tensor is $%
T_{00}=\rho$ and one may easily find that $R=-2\nabla^2 \Phi,$ where $\Phi$
is the Newtonian potential which is related to the metric component through $%
g_{00}=-1+2\Phi$. Note that in this paper we are considering the vector
field $A_\mu$ as the dark energy sector of the universe which should be
very small in the Newtonian limit of the theory. So, we consider $A_\mu$ as
a first order perturbed field, the same order as $\Phi$.

The trace of the equation \eqref{met} can be reduced to
\begin{align}
-R+\f{4\pi}{m^2}\bigg[\nabla_{\alpha}F_{\beta\mu}\nabla^\alpha F^{\beta\mu}
&+2\nabla_\alpha \big(F^{\alpha\beta}\nabla^\mu F_{\mu\beta} +F_{\beta\mu}\nabla^{\mu}F^{\alpha\beta}+F_{\beta\mu}\nabla^\alpha
F^{\mu\beta}\big)\bigg]  \notag \\
&+16\pi \bigg(A^2 V^\prime(A^2) -2V(A^2)\bigg)-24\pi\beta A_\alpha u^\alpha
p=8\pi T,
\end{align}
where $T$ is the trace of the matter energy momentum tensor. We have to keep only the first order terms in $\Phi$ and $A_\mu$. This
implies that the terms in the metric equation \eqref{met} which are
quadratic in $A_\mu$ do not contribute to this limit. With these assumptions
one  obtains the generalized Poisson equation
\begin{align}
\nabla^2 \Phi=-4\pi \rho -8\pi (A^2 V^\prime(A^2) - 2V(A^2))|_{\mathcal{O}(1)},
\end{align}
where $\mathcal{O}(1)$ means that we only keep terms which are linear in $%
A_\mu$. This implies that only the $V(A^2)=const.$  affect the Poisson
equation, which is exactly the cosmological constant. Note that the minus sign behind the energy density is because of our convention in defining the Newtonian potential in $g_{00}$.

Let us now consider the vector field equation \eqref{vec} in the Newtonian
limit. In this limit, the covariant derivatives should be replaced by
partial derivatives, since we have assumed that the vector field $A_\mu$ is
a small quantity. One can then show that the vector field equation reduces
to
\begin{align}
\partial_\mu F^{\mu\nu}+\f{1}{m^2}\partial_\mu \partial^\alpha \partial_\alpha
F^{\mu\nu}=\beta J^\nu,
\end{align}
which is exactly the original Bopp-Podolsky equation.

\section{Cosmological implications in the presence of Bopp-Podolsky type vector fields}\label{sect3}
\subsection{The Isotropic Cosmology}
In this section we want to consider the cosmological implications of the theory. First, let us assume that the geometry of the Universe is described by the Fiedmann-Robertson-Walker metric. With this choice the possible form of the vector field $A_\mu$ should have the form
$$A_\mu=(A_0(t),0,0,0)$$
to preserve homogeneity and isotropy. However, with this choice the vector field strength tensor $F_{\mu\nu}$ and therefore the Bopp-Podolsky term vanishes in our theory. One can easily find that the Friedmann and the vector field equations in the absence of matter fields in this case can be written as
\begin{align}
&3H^2+8\pi V(A_0^2)+16\pi A_0^2 V^\prime(A_0^2)=0,\\
&3H^2+2\dot{H}+8\pi V(A_0^2)=0,\\
&V^\prime(A_0^2)=0.
\end{align}
The simplest possibility to satisfy the last equation is that the potential becomes constant. This is the standard de Sitter type theory, with constant Hubble parameter
$H=H_0=\sqrt{8\pi V/3}$. One can however drive a self accelerated expanding universe by choosing other forms for the potential $V(A^2)$. In these cases the $(0)$-component of the vector field should be constant in order to satisfy the equation $V^\prime(A_0^2)$. For example, in the case that $V(A^2)=\alpha A^2+\beta A^4$, one should have $A_0=\sqrt{\alpha/2\beta}$.
\subsection{Anisotropic Cosmology - Bianchi I model}
In order to make the theory non-trivial, we should assume that the vector field has a spatial component. So, we will assume that the Universe can be described by the Bianchi-I type metric of the form
\begin{align}
ds^2=-dt^2+a^2(t)dx^2+b^2(t)\big(dy^2+dz^2\big),
\end{align}
and the vector field can then be written as
\begin{equation}
A_\mu=\big(0,B(t),0,0\big).
\end{equation}
One should note that we have assumed that the $(0)$ component of the vector field is zero. This is because this component does not contribute to the strength tensor $F_{\mu\nu}$.

Also, we assume that the matter content of the Universe consists of a
perfect cosmological fluid, with energy momentum tensor
\begin{align}
T^\mu_{~\nu}=\mathrm{diag}(-\rho(t),p(t),p(t),p(t)),
\end{align}
where $\rho $ is the total matter density (dark plus baryonic), and $p$ is
the matter thermodynamic pressure.

For later convenience, we will define the directional Hubble factors $H_i$, the mean Hubble factor $H$, the anisotropy parameter $A$, the shear scalar $\bar{\Sigma}^2$ and the deceleration parameter $q$ as \cite{def}
\begin{align}
H_1&=\f{\dot{a}}{a},\qquad H_2=\f{\dot{b}}{b},\\
H&=\f{1}{3}\sum_{i=1}^{3} H_i=\f{1}{3}(H_1+2H_2),\\
A&=\f{1}{3}\sum_{i=1}^{3}\left(\f{\Delta H_i}{H}\right)^2,\quad\textmd{with}\quad\Delta H_i=H-H_i,\\
\bar{\Sigma}^2&=\f{1}{2}\left(\sum_{i=1}^{3} H_i^2-3H^2\right)=\f{3}{2}AH^2=\f{3}{4}(H-H_1)^2,\\
q&=\frac{d}{dt}\left(\frac{1}{H}\right)-1.
\end{align}

With the above definitions, one can easily see that the quantity $\nabla_\mu(V^\prime A^\mu)$ vanishes. In this case, the time component of the conservation equation \eqref{conE}  leads to the usual conservation equation of the form
\begin{align}
\dot{\rho}+(H_1+2H_2)(\rho+p)=0,
\end{align}
while the $(x)$-component of the conservation equation gives $\beta\rho\dot{B}=0$. We can then assume that $B={\rm constant}$, a condition  which further implies that the Bopp-Podolsky term vanishes, or one should conclude that $\beta=0$, i.e. no matter/vector field coupling. We will choose the second choice and in the following we will assume that $\beta=0$ and then the conservation equation for the ordinary matter field hold. With these assumptions, one can see that equation \eqref{sh1} is satisfied identically.

\subsubsection{The cosmological field equations}

Let us introduce a new variable $F$, defined as
\begin{equation}
F=\frac{\dot{B}}{a}.
\end{equation}
With the above assumptions, only the $(x)$ component of the vector field equation of motion becomes non-zero, which can be written as
\begin{align}
\dddot{F}+&(5H+\Sigma)\ddot{F}+\f12\Sigma(2H-\Sigma)\dot{F}+(4H^2+3\dot{H})\dot{F}-m^2\dot{F}\nonumber\\&-(\Sigma+2H)(\dot{\Sigma}+2\dot{H}+3H\Sigma+m^2)F-\f12(8H^3+\Sigma^3)F+2m^2\f{B}{a}V^\prime(t)=0.
\end{align}
The Friedmann equations can then be simplified to
\begin{align}
-3H^2+\f34\Sigma^2-\f{6\pi}{m^2}(\Sigma^2+4H^2+4H\Sigma)F^2-\f{4\pi}{m^2}(\dot{F}-6HF)\dot{F}+\f{4\pi}{m^2}(2\ddot{F}+m^2F)F-8\pi V+8\pi\rho=0,
\end{align}
\begin{align}
-2\dot{H}-\dot{\Sigma}-3H^2-3H\Sigma-&\f34\Sigma^2-\f{2\pi}{m^2}(\Sigma^2+4H\Sigma+4H^2)F^2+\f{4\pi}{m^2}(\dot{F}+6HF)\dot{F}\nonumber\\&+\f{4\pi}{m^2}(2\ddot{F}+m^2F)F+8\pi\left(2\f{B^2}{a^2}V^\prime-V\right)-8\pi p=0,
\end{align}
and
\begin{align}
-2\dot{H}+\f12\dot{\Sigma}-&3H^2-\f34\Sigma^2+\f32H\Sigma+\f{4\pi}{m^2}(\dot{F}-2F\Sigma)\dot{F}+\f{2\pi}{m^2}(\Sigma^2-8H^2)F^2\nonumber\\&-\f{4\pi}{m^2}(2\dot{H}+\dot{\Sigma})F^2-\f{4\pi}{m^2}(4\dot{F}+F\Sigma)HF-4\pi F^2-8\pi V-8\pi p=0,
\end{align}
where we have defined $\Sigma=2/\sqrt{3}\bar{\Sigma}$.
\subsubsection{The case of massless vector field}

Let us now investigate  the cosmological implications of the Bopp-Podolsky theory with a massless vector field. In this case the potential term $V(A^2)$ vanishes.
In order to simplify the mathematical formalism, let us introduce a set of
dimensionless variables $\left( \tau ,r,P,f,h,\mu,\sigma\right) $, defined as
\begin{eqnarray}  \label{var}
\tau =H_{0}t,\quad \rho =\frac{3H_{0}^{2}}{8\pi }r,\quad p=\frac{3H_{0}^{2}}{8\pi }P
,\quad m=\mu H_0,\quad F =H_{0}f,\quad H=H_{0}h,\quad \Sigma=\sigma H_0,
\end{eqnarray}%
where $H_{0}$ is the present day value of the Hubble function. By using the
above set of variables, the cosmological evolution equations for the
Bopp-Podolsky type vector dark energy model can be written as
\begin{align}\label{54}
\dddot{f}+&(5h+\sigma)\ddot{f}+\f12\sigma(2h-\sigma)\dot{f}+(4h^2+3\dot{h})\dot{f}-\mu^2\dot{f}-(\sigma+2h)(\dot{\sigma}+2\dot{h}+3h\sigma+\mu^2)f-\f12(8h^3+\sigma^3)f=0,
\end{align}
\begin{align}\label{55}
-3h^2+\f34\sigma^2-\f{6\pi}{\mu^2}(\sigma^2+4h^2+4h\sigma)f^2-\f{4\pi}{\mu^2}(\dot{f}-6hf)\dot{f}+\f{4\pi}{\mu^2}(2\ddot{f}+\mu^2f)f+3r=0,
\end{align}
\begin{align}\label{56}
-2\dot{h}-\dot{\sigma}-3h^2-3h\sigma-&\f34\sigma^2-\f{2\pi}{\mu^2}(\sigma^2+4h\sigma+4h^2)f^2+\f{4\pi}{\mu^2}(\dot{f}+6hf)\dot{f}+\f{4\pi}{\mu^2}(2\ddot{f}+\mu^2f)f-3P=0,
\end{align}
and
\begin{align}\label{57}
-2\dot{h}+\f12\dot{\sigma}-&3h^2-\f34\sigma^2+\f32h\sigma+\f{4\pi}{\mu^2}(\dot{f}-2f\sigma)\dot{f}+\f{2\pi}{\mu^2}(\sigma^2-8h^2)f^2\nonumber\\&-\f{4\pi}{\mu^2}(2\dot{h}+\dot{\sigma})f^2-\f{4\pi}{\mu^2}(4\dot{f}+f\sigma)hf-4\pi f^2-3P=0,
\end{align}
and the energy conservation equation becomes
\begin{equation}
\dot{r}+3h\left(r+P\right)=0,
\end{equation}
where now, ``dot" represents derivative with respect to $\tau$. One should note that because we want to make the ghost degree of freedom non-dynamical, one should assume $\mu\gg1$.

From Eq.~(\ref{55}) we can obtain $\ddot{f}$ as
\begin{align}
\ddot{f}=\frac{16 \pi  \dot{f}\left(\dot{f}-6fh\right)+8 \pi  f^2 \left[3 (2 h+\sigma
   )^2-2 \mu ^2\right]+3 \mu ^2 \left(4 h^2-4 r-\sigma ^2\right)}{32 \pi
   f}.
\end{align}
After substituting this expression of $\ddot{f}$ into Eq.~(\ref{56}), we can solve Eqs.~(\ref{56}) and (\ref{57}) to obtain the expressions of $\dot{h}$ and $\dot{\sigma}$, respectively. Therefore the system of equations describing the evolution of the anisotropic Bianchi type I Universe in the presence of Bopp-Podolsky type vector dark energy can be written as
\be\label{60}
\dot{f}=u,
\ee
\begin{align}\label{61}
\dot{u}=\frac{16 \pi  u\left(u-6fh\right)+8 \pi  f^2 \left[3 (2 h+\sigma
   )^2-2 \mu ^2\right]+3 \mu ^2 \left(4 h^2-4 r-\sigma ^2\right)}{32 \pi
   f},
\end{align}
\bea\label{62}
\dot{\sigma}&=&\frac{1}{6 \mu ^4}\Bigg\{16 \pi  f^2 \left[\mu ^2 ((4 h-\sigma ) (2 h+\sigma )-3 (P+r))+\mu ^4+8 \pi  u^2\right]+64 \pi ^2 f^4 (2 h+\sigma )^2+32 \pi
   \mu ^2 u f (2 h+\sigma )+\nonumber\\
&&   3 \mu ^4 \left(-6 h \sigma +4 h^2-4 r-\sigma ^2\right)+16 \pi  \mu ^2 u^2\Bigg\},
\eea
\bea\label{63}
\dot{h}&=&\frac{1}{6 \mu ^4}\Bigg\{4 \pi  f^2 \left[\mu ^2 \left(8 h \sigma -4 h^2+6 (P+r)+5 \sigma ^2\right)-2 \left(\mu ^4+8 \pi  u^2\right)\right]-32 \pi ^2 f^4 (2
   h+\sigma )^2-16 \pi  \mu ^2 f u (2 h+\sigma )-\nonumber\\
 &&  3 \mu ^4 \left(2 h^2+3 P+r+\sigma ^2\right)+16 \pi  \mu ^2 u^2\Bigg\},
\eea
\be\label{64}
\dot{r}=-3h(r+P),
\ee
\be\label{65}
P=P(r).
\ee

After adopting an equation of state for the cosmological matter, the system
of differential equations Eqs.~(\ref{60})-(\ref{64}) must be solved by
choosing some appropriate initial conditions, which we take as $f(0)=f_0$, $u(0)=u_0$, $%
h(0)=h_0$, $\sigma(0)=\sigma _0$, and $r(0)=r_0$, respectively. In the dimensionless variables introduced above the
deceleration parameter is given by
\begin{equation}
q=\frac{d}{dt}\left(\frac{1}{h}\right)-1.
\end{equation}

In the following we will assume that the age of the Universe is of the order of $t_{max}=(2/3)t_H$, which gives for the dimensionless time $\tau $ the maximum value of $\tau _{max}=2/3=0.66$.

\subsubsection{Approximate anisotropic solution with constant $f$}\label{3C}

As an example of a simple exact solution of the cosmological evolution equation (\ref{54})-(\ref{57}) we will consider the case $f=f_0={\rm constant}$, a condition which gives for the evolution of the vector field potential $B$ an equation of the form $\dot{B}=f_0a$. With this choice for $f$ the evolution equations (\ref{54})-(\ref{57}) become
\be\label{64a}
-\frac{1}{2} f_0(2 h+\sigma ) \left[2 \left(2 \dot{h}+\mu ^2+\dot{\sigma} \right)+\left(2 h+\sigma \right)^2\right]=0,
\ee
\be\label{65a}
-\frac{6 \pi  f_0^2 \left(2 h+\sigma \right)^2}{\mu ^2}+4 \pi  f_0^2-3 h^2+3 r+\frac{3 \sigma ^2}{4}=0,
\ee
\be\label{66a}
-\frac{\left(8 \pi  f_0^2+3 \mu ^2\right) \left(2 h+\sigma \right)^2}{4 \mu ^2}+4 \pi  f_0^2-2 \dot{h}-\dot{\sigma} -3 P=0,
\ee
\be\label{67a}
8 \pi f_0^2 \left[\sigma ^2-2 \left(2 \dot{h}+\mu ^2+\dot{\sigma}\right)\right]+2 \left(3 \mu ^2-8 \pi  f_0^2\right) h \sigma -4 \left(16 \pi  f_0^2+3
   \mu ^2\right) h^2-\mu ^2 \left(8 \dot{h}+12 P-2 \dot{\sigma} +3 \sigma ^2\right)=0.
\ee

Eqs.~(\ref{65a})-(\ref{67a}) can be solved to give the matter energy density and the pressure as
\be
r= \frac{(2 h+\sigma ) \left[2 \left(8 \pi  f_0^2+\mu ^2\right) h+\left(8 \pi  f_0^2-\mu ^2\right) \sigma \right]}{4 \mu ^2}-\frac{4
   \pi  f_0^2}{3},
\ee
\be\label{69}
P= \frac{1}{3} \left(-\frac{\left(8 \pi  f_0^2+3 \mu ^2\right) \left(2 h+\sigma \right)^2}{4 \mu ^2}+4 \pi  f_0^2-2 \dot{h}-\dot{\sigma }
   \right),
\ee
\be\label{70}
P= \frac{-8 \left(4 \pi  f_0^2+\mu ^2\right) \dot{h}+2 \left(3 \mu ^2-8 \pi  f_0^2\right) h \sigma -4 \left(16 \pi  f_0^2+3 \mu
   ^2\right) h^2-16 \pi  f_0^2 \mu ^2+2 \left(\mu ^2-8 \pi  f_0^2\right) \dot{\sigma }+\left(8 \pi  f_0^2-3 \mu ^2\right) \sigma ^2}{12 \mu
   ^2}.
\ee
By assuming $2h+\sigma \neq 0$, Eq.~(\ref{64a}) can be immediately integrated to give
\be
\sigma =-2
   h -\sqrt{2} \mu  \tan \left[\frac{\mu  \left(\tau-\tau_0\right)}{\sqrt{2}}-\delta\right],
\ee
where we have denoted
\be
\delta =\tan ^{-1}\left(\frac{2 h_0+\sigma _0}{\sqrt{2} \mu }\right),
\ee
and we have used the initial condition $h\left(\tau _0\right)=h_0$ and $\sigma \left(\tau _0\right)=\sigma _0$, respectively. Then the requirement of the equality of the pressures in Eqs.~(\ref{69}) and (\ref{70}) gives for $h$ the evolution equation
\begin{equation}\label{73a}
\dot{h}+3\frac{\sqrt{2}\left( 8\pi f_{0}^{2}-3\mu ^{2}\right) }{6\mu }\tan %
\left[ \delta +\frac{\mu (\tau _{0}-\tau )}{\sqrt{2}}\right] h+3h^{2}+\frac{1%
}{2}\left( \mu ^{2}-8\pi f_{0}^{2}\right) \sec ^{2}\left[ \delta +\frac{\mu
(\tau _{0}-\tau )}{\sqrt{2}}\right] +\frac{16}{3}\pi f_{0}^{2}=0.
\end{equation}

Eq.~(\ref{73a}) is a Riccati type equation, which generally cannot be solved exactly. For the matter energy density and pressure we obtain
\be
r= \left(4 \pi  f_0^2-\frac{\mu ^2}{2}\right) \tan ^2\left[\delta +\frac{\mu  (\tau _0-\tau)}{\sqrt{2}}\right]+\sqrt{2} \mu
    h \tan \left[\delta +\frac{\mu  (\tau _0-\tau)}{\sqrt{2}}\right]-\frac{4 \pi  f_0^2}{3},
\ee
\be
P= \frac{1}{6} \left\{16 \pi  f_0^2+3 \mu
   ^2-\left(8 \pi  f_0^2+\mu ^2\right) \sec ^2\left[\delta +\frac{\mu  (\tau _0-\tau)}{\sqrt{2}}\right]\right\}.
\ee
The solutions are periodic, with the period $T=2\sqrt{2}\pi /\left(\mu H_0\right)=2\sqrt{2}\pi t_H/ \mu$, where $t_H=1/H_0$ is the present day age of the Universe. Hence one period describes roughly the entire cosmological history. In the rescaled dimensionless time $\tau $ this corresponds to a time interval $2\sqrt{2}\pi /\mu $.

An approximate simple solutions of Eq.~(\ref{73a}) can be obtained by assuming the conditions $\mu ^2=8\pi f_0^2$, and $\mu \left(\tau -\tau _0 \right)/\sqrt{2}<<\delta$. Then Eq.~(\ref{73a}) takes the form
\be
\dot{h}-\left(2h_0+\sigma _0\right)h+3h^2+\frac{2}{3}\mu ^2=0,
\ee
with the general solution
\be
h= \frac{1}{6} \left\{\sqrt{\left(2 h_0+\sigma _0\right)^2-8 \mu
   ^2} \tanh \left[\frac{1}{2} \sqrt{\left(2 h_0+\sigma _0\right)^2-8 \mu ^2}\left(\tau-3 c_1\right) \right]+2 h_0+\sigma _0\right\},
\ee
where $c_1$ is an arbitrary constant of integration, and we have assumed $\left(2 h_0+\sigma _0\right)^2>8 \mu ^2$. In the limit of large times the mean Hubble function tends to a constant, $\lim_{\tau \rightarrow \infty}h=\left[\sqrt{\left(2 h _0+\sigma _0\right)^2-8 \mu ^2}+2 h _0+\sigma _0\right]/6={\rm constant}$,  while the volume $\cal{V}$ of the Universe increases according to
\be
\mathcal{V} =ab^2=e^{\frac{1}{2}\left(2 h_0+\sigma _0\right) t } \cosh \left[\frac{1}{2}\sqrt{\left(2 h_0+\sigma _0\right)^2-8 \mu ^2} \left(\tau-3 c_1\right) \right].
\ee
For the deceleration parameter we obtain
\be
q=-\frac{3 \left[\left(2 h _0+\sigma _0\right)^2-8 \mu ^2\right] \text{sech}^2\left[\frac{1}{2}\sqrt{\left(2 h _0+\sigma _0\right)^2-8 \mu
   ^2} \left(\tau-3 c_1\right) \right]}{\left\{\sqrt{(2 h _0+\sigma _0)^2-8 \mu ^2} \tanh \left[\frac{1}{2}\sqrt{\left(2 h _0+\sigma _0\right)^2-8 \mu ^2} \left(\tau-3 c_1\right) \right]+2
   h _0+\sigma _0\right\}{}^2}-1
\ee
In the large time limit $\lim_{\tau \rightarrow \infty}q=-1$, and thus the anisotropic Universe ends in a de Sitter exponentially accelerating phase. However, super-accelerating phases of evolution with $q<-1$ are also possible. In the same limit we obtain $\lim_{\tau \rightarrow \infty}\sigma =\sqrt{2}\mu -\left[\sqrt{\left(2 h _0+\sigma _0\right)^2-8 \mu ^2}+2 h _0+\sigma _0\right]/3$, giving for the mean anisotropy parameter
\be
\lim_{\tau \rightarrow \infty}A=\frac{8}{3}\frac{\left\{\sqrt{2}\mu -\left[\sqrt{\left(2 h _0+\sigma _0\right)^2-8 \mu ^2}+2 h _0+\sigma _0\right]\right\}^2}{\left\{\left[\sqrt{\left(2 h _0+\sigma _0\right)^2-8 \mu ^2}+2 h _0+\sigma _0\right]\right\}^2}.
\ee
In the large time limit the matter energy density and pressure also reach some constant values, strongly dependent on the model parameters.

\subsubsection{Cosmological evolution of the anisotropic radiation fluid Universe}

The radiation epoch, in which the Universe consisted of a plasma of nuclei, electrons and photons, is one of the most important periods in the evolution of the Universe. During this period the temperature was in the range of $10^9-10^3$ K, and the  temperatures remained too high for the binding of electrons to nuclei. Therefore during this phase the Universe was filled with a radiation fluid, described by the equation of state $P = r/3$. The radiation era lasted from around $t=10$ s to $t=10^{13}$ s \cite{Mukh},  giving for the dimensionless time $\tau $ the range $\tau \in \left[\tau_{in}=2.18\times 10^{-17},\tau _{fin}=2.18\times 10^{-5}\right]$, where for the present day value of the Hubble constant we have adopted the numerical value $H_0=67.31\;{\rm  km/Mpc\;s}=2.185\times 10^{-18}$ s \cite{Planck1}. We approximate the initial value of the Hubble function at the beginning of the radiation era as being given by $h\left(\tau _{in}\right)\approx 1/2\tau_{in}=2.28\times 10^{16}$, with the initial dimensionless density of the matter in the Universe given as $r\left(\tau _{in}\right)\approx 3h^2\left(\tau _{in}\right)=1.57\times 10^{33}$. The initial value of the shear scalar can be obtained as $\sigma \left(\tau_{in}\right)=\sqrt{(3/2)A\left(\tau_{in}\right)h^2\left(\tau _{in}\right)}$. We define (arbitrarily) the initial value of the anisotropy parameter as $A\left(\tau_{in}\right)=1$, which gives $\sigma \left(\tau_{in}\right)=\sqrt{(3/2)}h\left(\tau _{in}\right)$

\begin{figure*}[htp]
\begin{center}
\includegraphics[width=8.0cm]{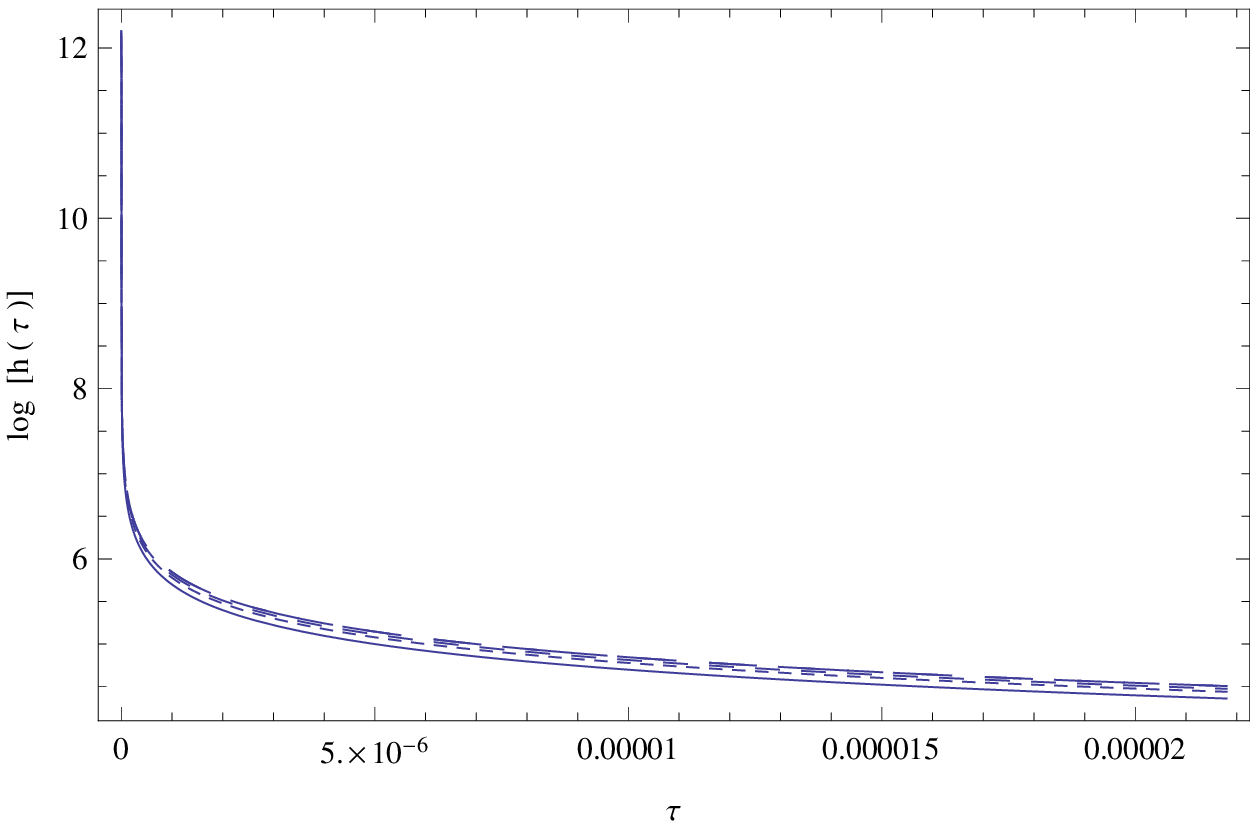}
\includegraphics[width=8.0cm]{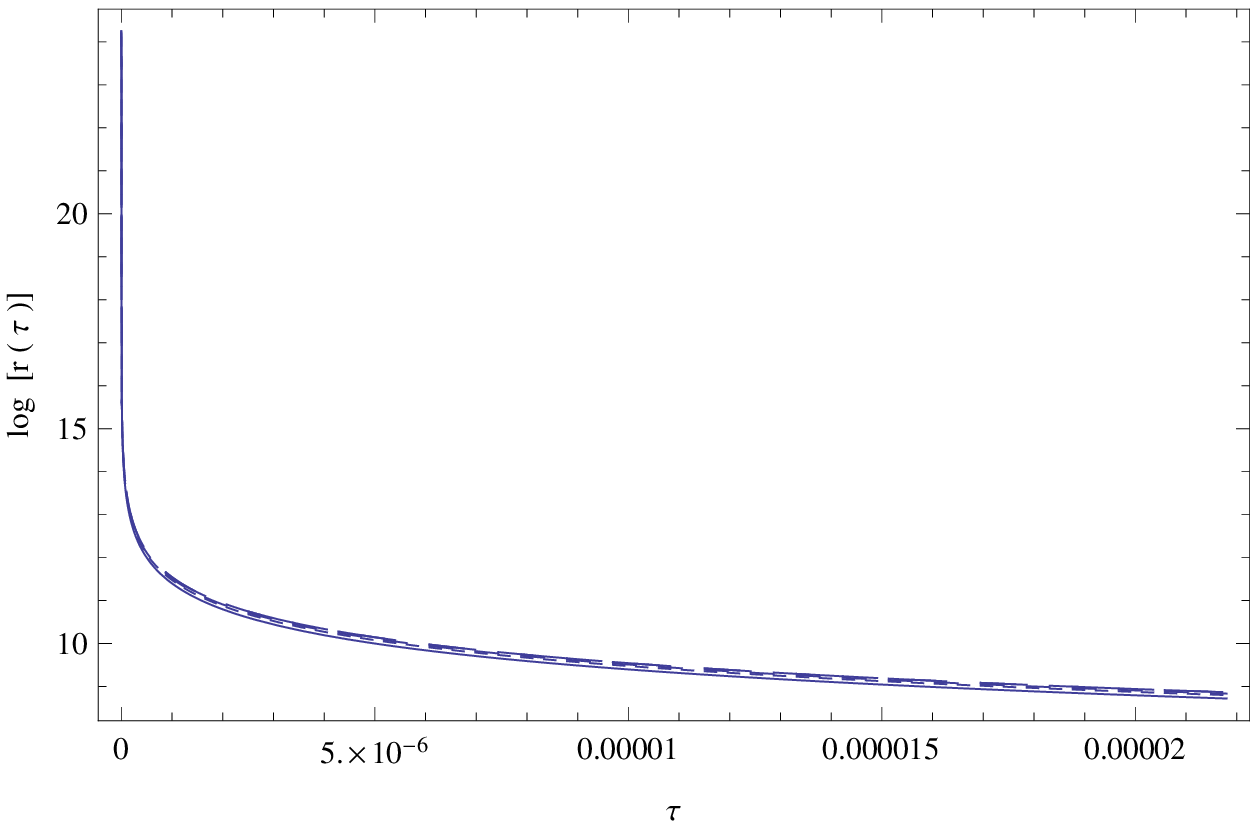}
\caption{Variation in a logarithmic scale of the Hubble function of the anisotropic radiation fluid Universe with
Bopp-Podolsky type vector dark energy (left figure), and of the matter energy density (right figure), for different values of the dimensionless Bopp-Podolsky parameter $\mu$: $\mu =160$ (solid curve), $\mu=140$ (dotted curve), $\mu=120 $ (short dashed curve), $\mu =100$ (dashed curve), and $\mu =80$ (long dashed curve), respectively. The initial conditions used to numerically integrate the cosmological evolution equations are $f\left(\tau _{in}\right)=2.55$, $u\left(\tau _{in}\right)=-0.001$, $\sigma \left(\tau _{in}\right)=2.80\times 10^{16}$, $h\left(\tau _{in}\right)=2.28\times 10^{33}$, and $r\left(\tau _{in}\right)=1.57\times 10^{33}$, respectively. }\label{fig1}
\end{center}
\end{figure*}

\begin{figure*}[htp]
\begin{center}
\includegraphics[width=8.0cm]{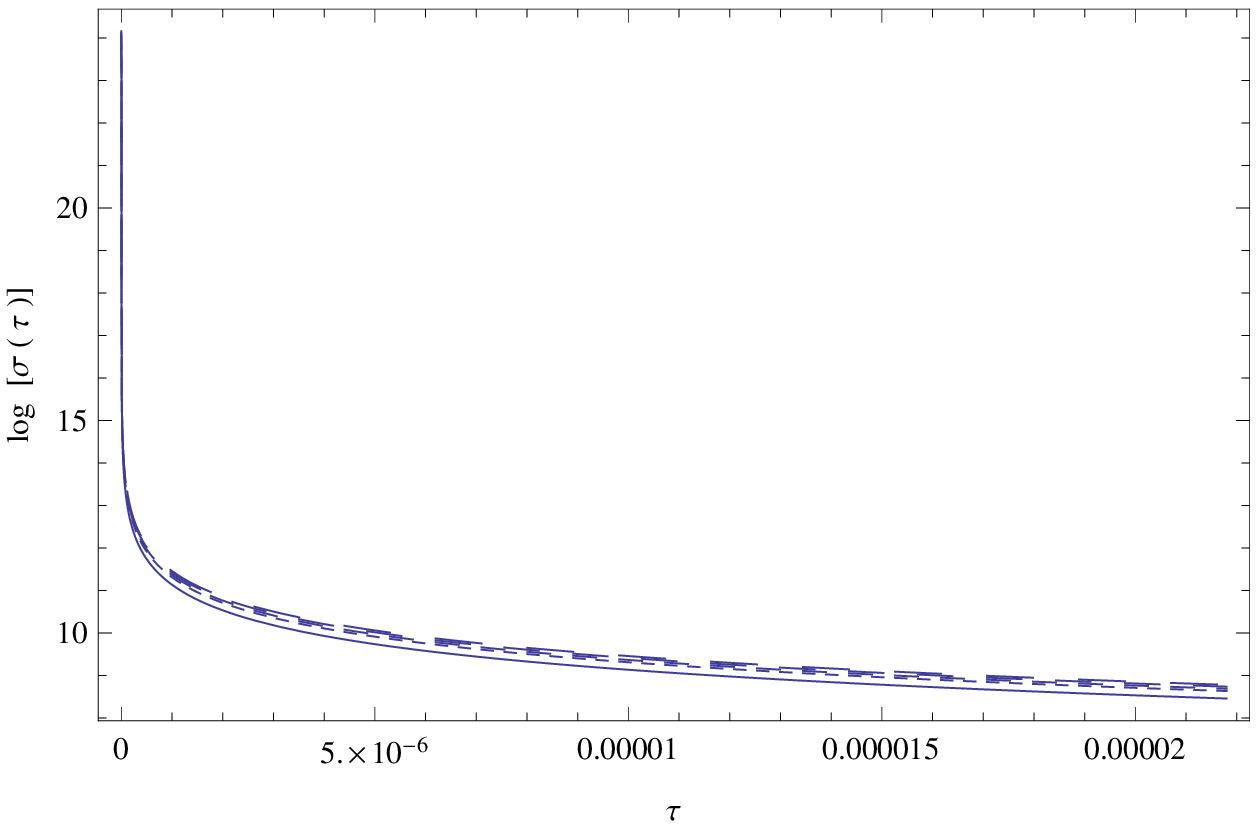}
\includegraphics[width=8.0cm]{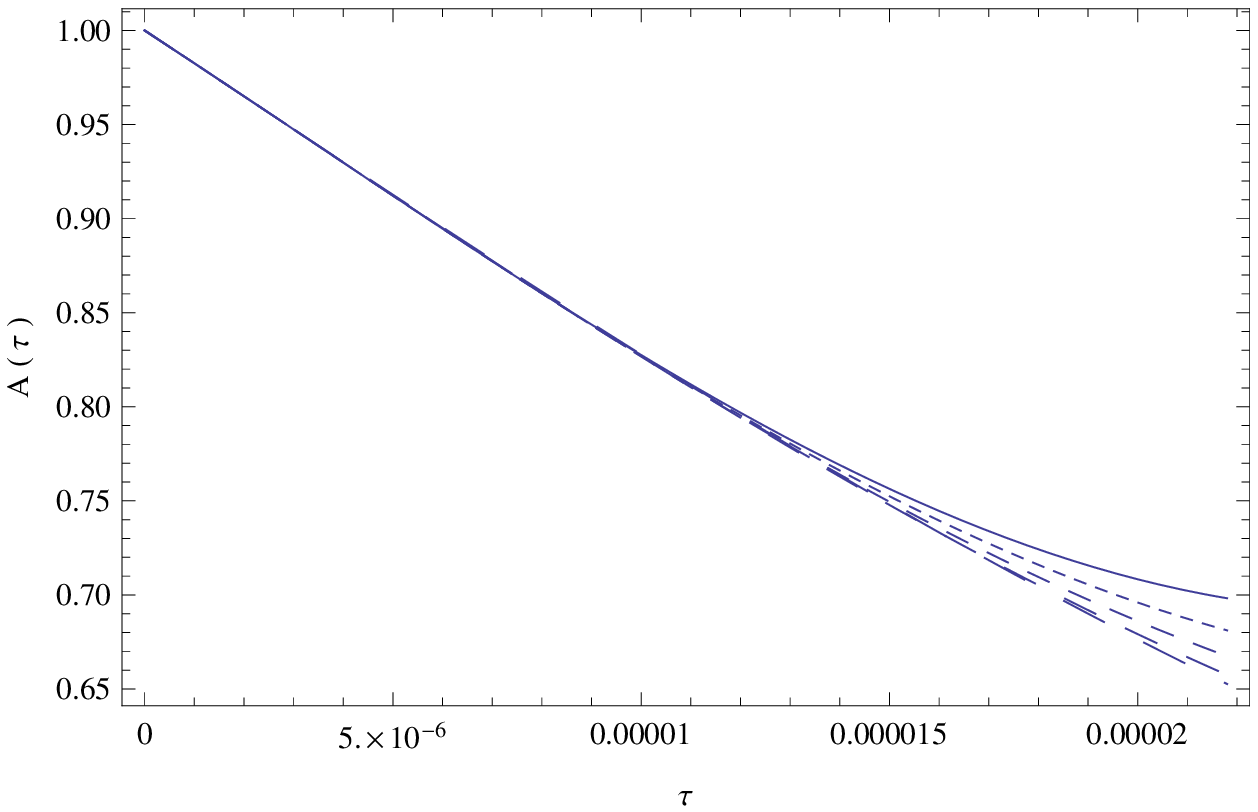}
\caption{Time evolution in a logarithmic scale of the shear scalar $\sigma $ (left figure) and of the anisotropy parameter $A$ (right figure) of the anisotropic radiation fluid Universe with Bopp-Podolsky type vector dark energy  for different values of the dimensionless Bopp-Podolsky parameter $\mu$: $\mu =160$ (solid curve), $\mu=140$ (dotted curve), $\mu=120 $ (short dashed curve), $\mu =100$ (dashed curve), and $\mu =80$ (long dashed curve), respectively. The initial conditions used to numerically integrate the cosmological evolution equations are $f\left(\tau _{in}\right)=2.55$, $u\left(\tau _{in}\right)=-0.001$, $\sigma \left(\tau _{in}\right)=2.80\times 10^{16}$, $h\left(\tau _{in}\right)=2.28\times 10^{33}$, and $r\left(\tau _{in}\right)=1.57\times 10^{33}$, respectively.}\label{fig2}
\end{center}
\end{figure*}

\begin{figure*}[htp]
\begin{center}
\includegraphics[width=8.0cm]{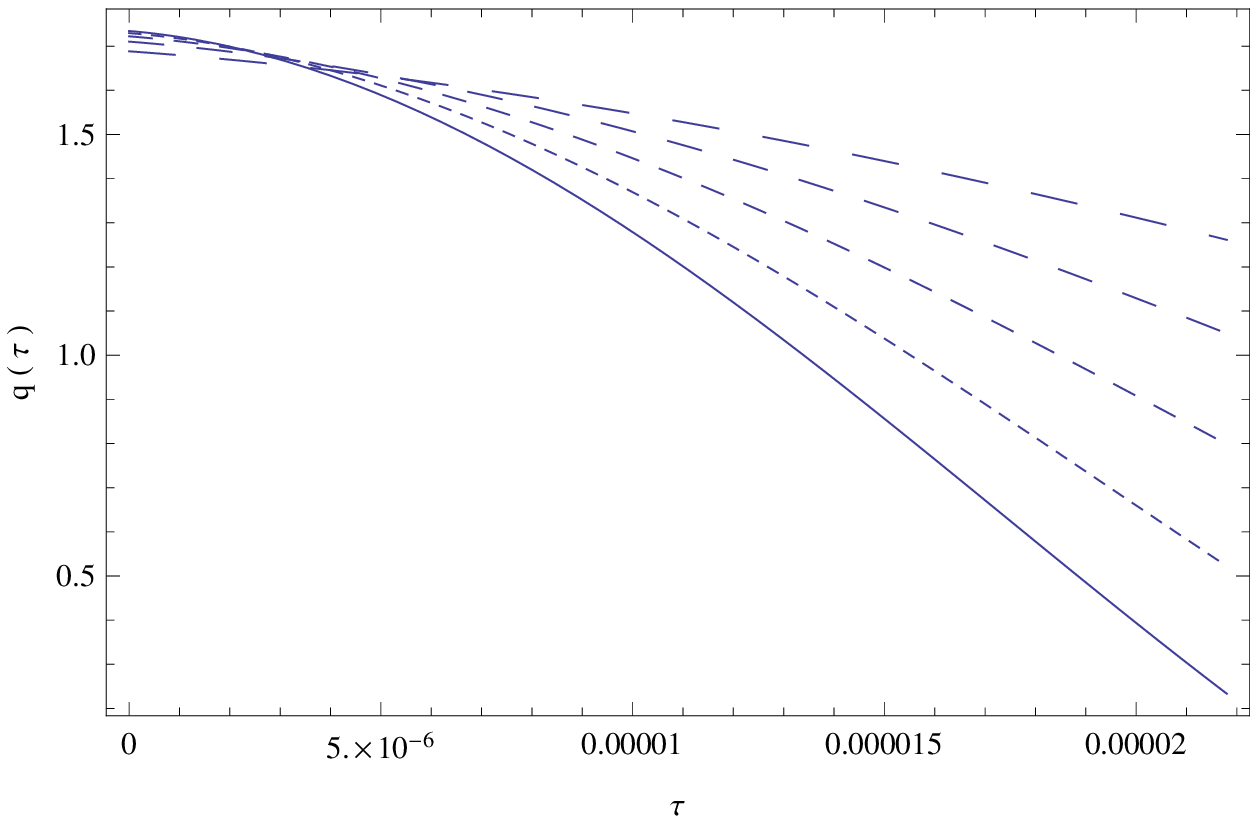}
\includegraphics[width=8.0cm]{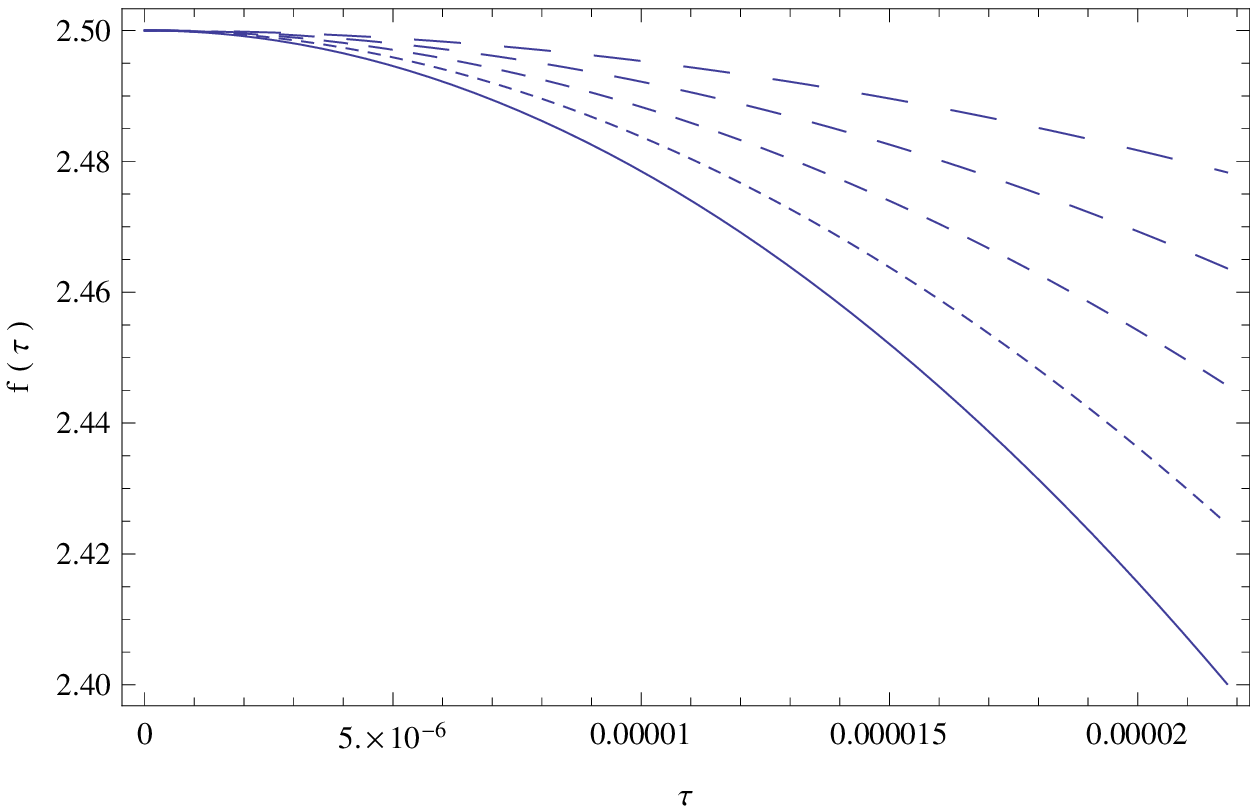}
\caption{Dynamics of the deceleration parameter $q$ (left figure) and of the Bopp-Podolsky vector potential (right figure) of the anisotropic radiation fluid Universe with Bopp-Podolsky type vector dark energy  for different values of the dimensionless Bopp-Podolsky parameter $\mu$: $\mu =160$ (solid curve), $\mu=140$ (dotted curve), $\mu=120 $ (short dashed curve), $\mu =100$ (dashed curve), and $\mu =80$ (long dashed curve), respectively.The initial conditions used to numerically integrate the cosmological evolution equations are $f\left(\tau _{in}\right)=2.55$, $u\left(\tau _{in}\right)=-0.001$, $\sigma \left(\tau _{in}\right)=2.80\times 10^{16}$, $h\left(\tau _{in}\right)=2.28\times 10^{33}$, and $r\left(\tau _{in}\right)=1.57\times 10^{33}$, respectively.}\label{fig3}
\end{center}
\end{figure*}

In order to obtain the evolution of the anisotropic Universe in the presence of a Bopp-Podolsky type dark energy, we have integrated numerically Eqs.~(\ref{60})-(\ref{64}) by using the following initial conditions: $f\left(\tau_{in}\right)=0.55$, $u\left(\tau _{in}\right)=-1$, $\sigma \left(\tau _{in}\right)=2.80\times 10^{16}$, $h\left(\tau _{in}\right)=2.28\times 10^{16}$, and $r\left(\tau _{in}\right)=1.57\times 10^{33}$, respectively. The time variations of the mean Hubble function, matter energy density, shear scalar, anisotropy parameter, deceleration parameter, and of the ratio of the time variation of the Bopp-Podolsky vector potential and scale factor, respectively, are presented in Figs.~\ref{fig1}-\ref{fig3}, for different values of the field mass term $\mu$.

As one can see from Fig.~\ref{fig1}, the mean Hubble function is a monotonically decreasing function of time, indicating an expansionary evolution of the anisotropic Bianchi type I cosmological model. The cosmological evolution rate is practically independent on the numerical values of $\mu $. In the large time limit very small differences in the expansion rate, determined by the variation of $\mu $, may appear. The radiation fluid  energy
density, shown in the right panel of Fig.~\ref{fig1}, is also a monotonically decreasing function of the cosmological time, and its time evolution is not affected significantly by the variations of the numerical values of $\mu$. The shear scalar $\sigma $, shown in the left panel of Fig.~\ref{fig2}, decreases  rapidly during the cosmological evolution, showing an almost linear dependence on $\tau $. The evolution of the shear scalar is influenced by the numerical values of $\mu$ only in the large time limit. A similar dynamics can be seen for the time evolution of the anisotropy parameter $A$, which decreases significantly, indicating the tendency of the anisotropic Universe to evolve towards an isotropic stage. The evolution of $A$ depends on the numerical values of $\mu $ only at the late phases of the radiation era. The mean deceleration parameter, presented in Fig.~\ref{fig3}, has only positive values, indicating a decelerating expansion, which essentially depends on the numerical values of $\mu $. The time variation of the Bopp-Podolsky vector potential $f$, depicted in the right panel of Fig.~\ref{fig3}, shows a strongly $\mu$-dependent dynamical evolution, with the function $f$ monotonically decreasing in time.

\subsubsection{Cosmological evolution of the anisotropic dust Universe}

As a second application of the anisotropic Bopp-Podolsky type cosmological model we consider the evolution of the dust, matter dominated, Bianchi type I Universe, with $P=0$. We assume that the matter dominated era began when the Universe was about 400,000 years old (after the recombination era), corresponding to an initial value of the dimensionless time coordinate of $\tau _{in}\approx 3\times 10^{-5}$. For the numerical value of the Hubble function at the beginning of the matter dominated era we adopt the value $h\left(\tau _{in}\right)=2.5\times 10^4$, while for the initial value of the dimensionless energy density of the matter we assume the value $r\left(\tau _{in}\right)=1.75\times 10^9$. We assume for the initial value of the anisotropy parameter at the beginning of the matter dominated era the value $A\left(\tau _{in}\right)=0.60$, giving for the initial value of the shear scalar $\sigma \left(\tau_{in}\right)=2.31\times 10^4$.

\begin{figure*}[htp]
\begin{center}
\includegraphics[width=8.0cm]{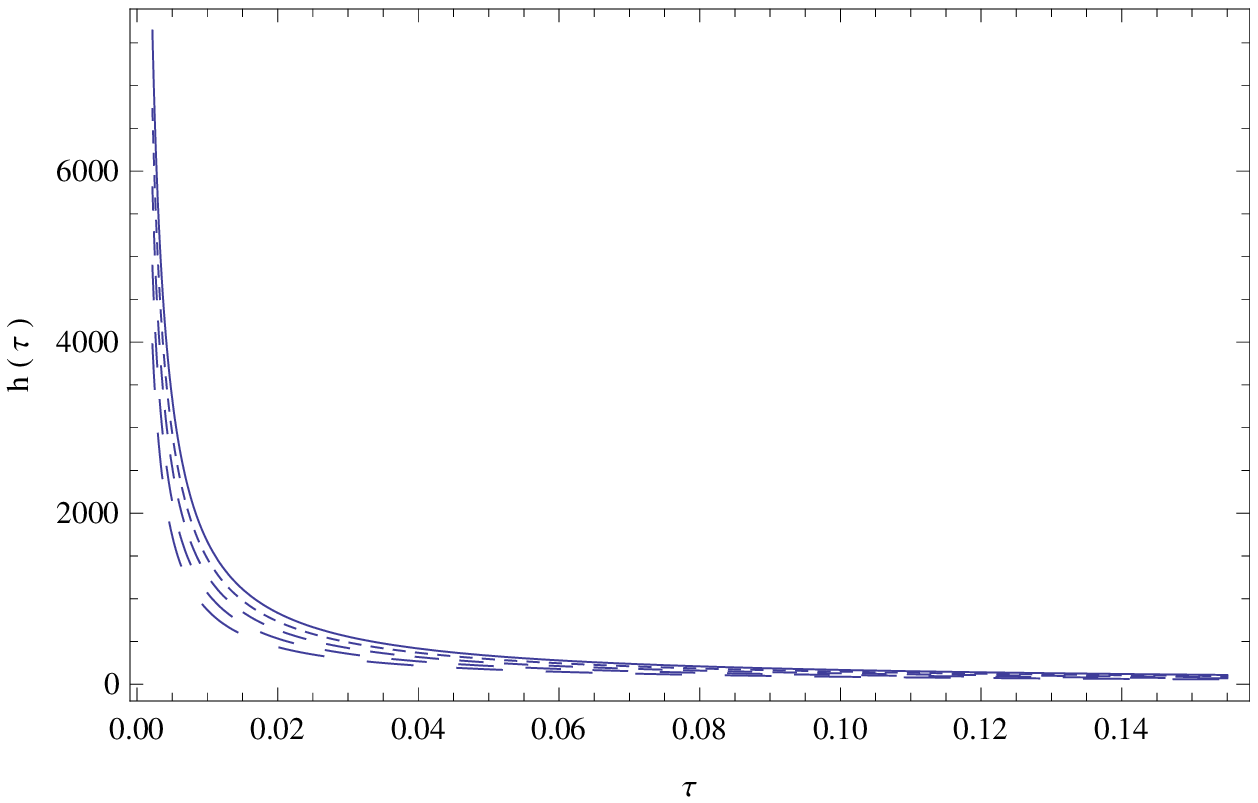}
\includegraphics[width=8.0cm]{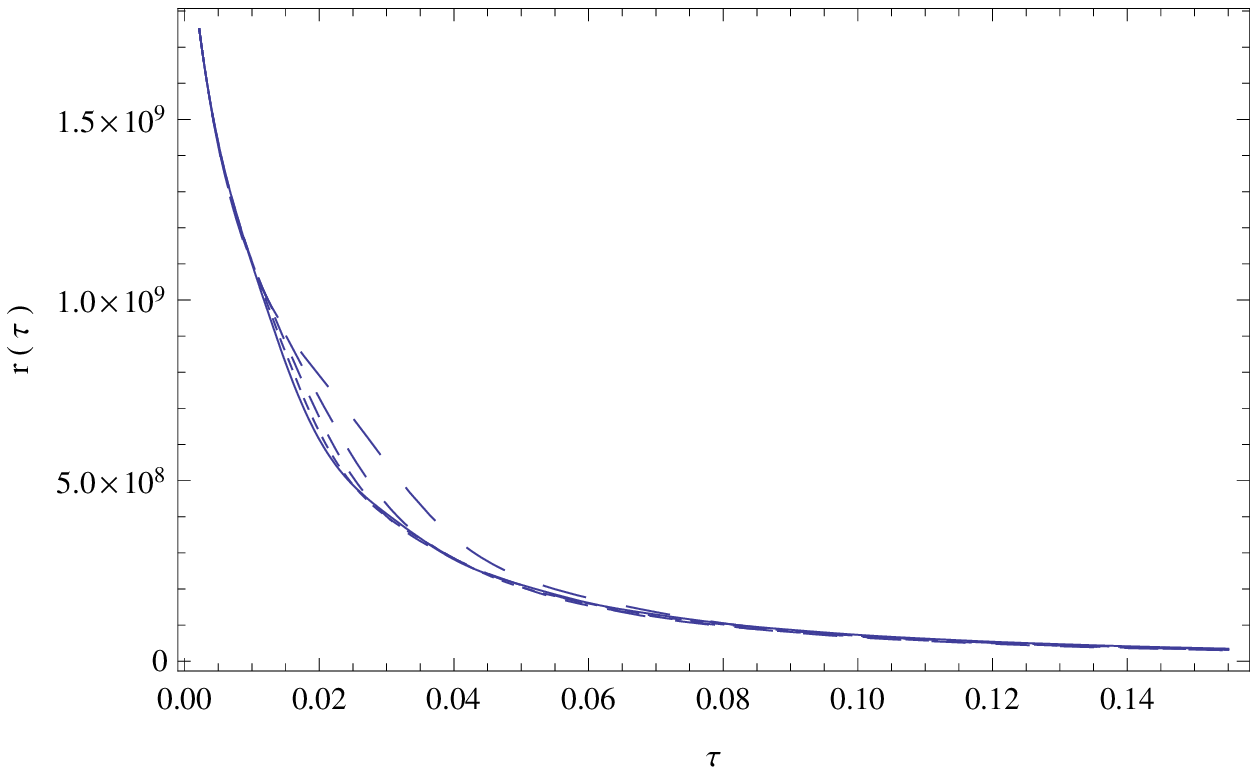}
\caption{Variation of the Hubble function of the anisotropic Bianchi type I dust Universe with
Bopp-Podolsky type vector dark energy (left figure), and of the matter energy density (right figure), for $\tau \geq \tau _{in}=3\times 10^{-5}$, and for different values of the dimensionless Bopp-Podolsky parameter $\mu$: $\mu =160$ (solid curve), $\mu=140$ (dotted curve), $\mu=120 $ (short dashed curve), $\mu =100$ (dashed curve), and $\mu =80$ (long dashed curve), respectively. The initial conditions used to numerically integrate the cosmological evolution equations are $f\left(\tau_{in}\right)=2.4$, $u\left(\tau_{in}\right)=-0.001$, $\sigma \left(\tau_{in}\right)=2.31\times 10^4$, $h\left(\tau _{in}\right)=2.5\times 10^4$, and $r\left(\tau _{in}\right)=1.75\times 10^9$, respectively.}\label{fig4}
\end{center}
\end{figure*}

\begin{figure*}[htp]
\begin{center}
\includegraphics[width=8.0cm]{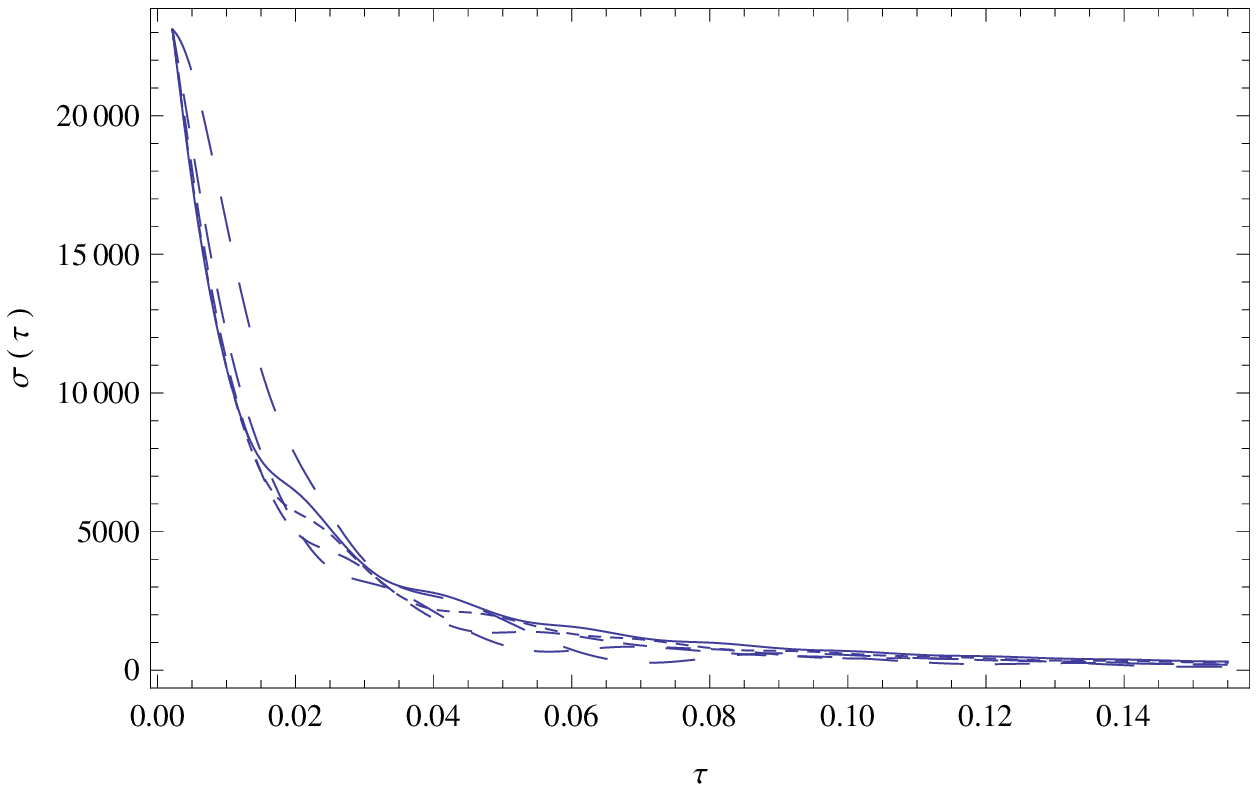}
\includegraphics[width=8.0cm]{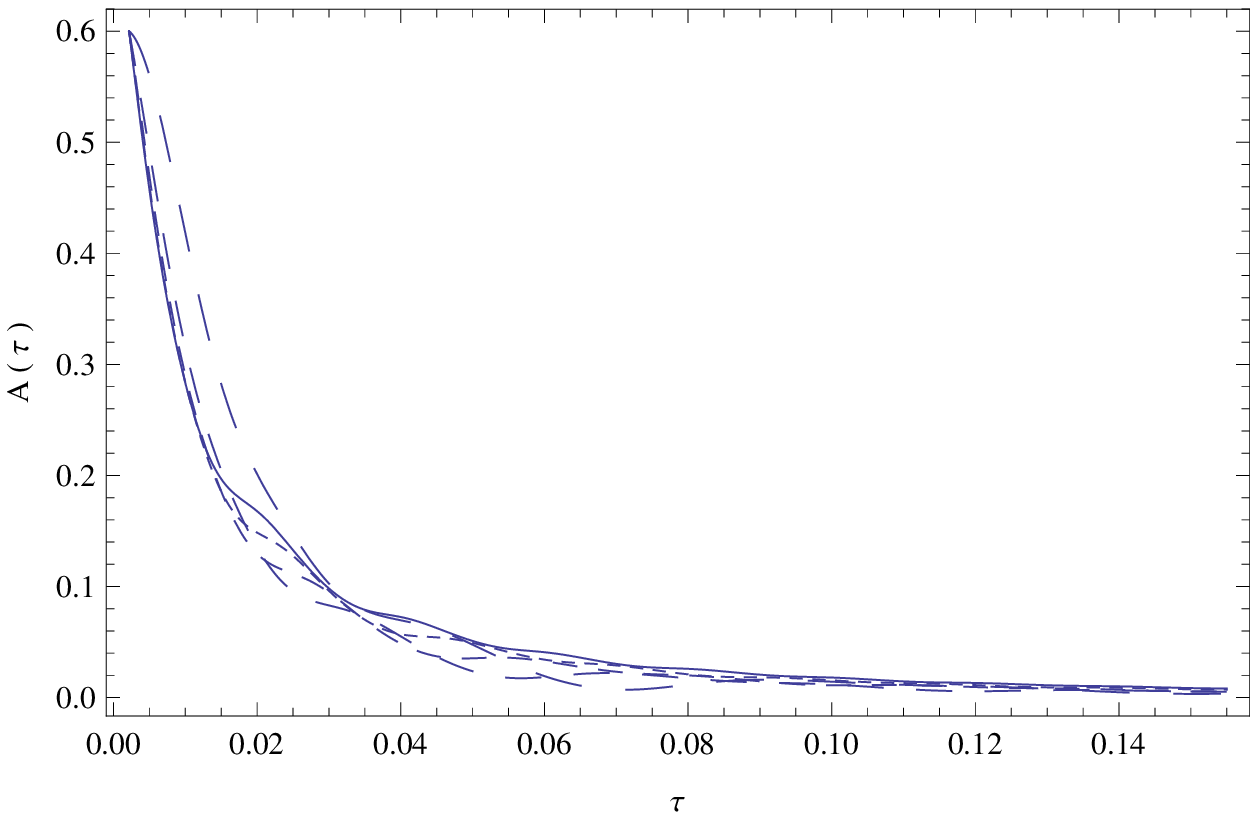}
\caption{Time evolution of the shear scalar $\sigma $ (left figure) and of the anisotropy parameter $A$ (right figure) of the anisotropic dust Universe with Bopp-Podolsky type vector dark energy  for $\tau \geq \tau _{in}=3\times 10^{-5}$, and for different values of the dimensionless Bopp-Podolsky parameter $\mu$: $\mu =160$ (solid curve), $\mu=140$ (dotted curve), $\mu=120 $ (short dashed curve), $\mu =100$ (dashed curve), and $\mu =80$ (long dashed curve), respectively. The initial conditions used to numerically integrate the cosmological evolution equations are $f\left(\tau_{in}\right)=2.4$, $u\left(\tau_{in}\right)=-0.001$, $\sigma \left(\tau_{in}\right)=2.31\times 10^4$, $h\left(\tau _{in}\right)=2.5\times 10^4$, and $r\left(\tau _{in}\right)=1.75\times 10^9$, respectively.}\label{fig5}
\end{center}
\end{figure*}

\begin{figure*}[htp]
\begin{center}
\includegraphics[width=8.0cm]{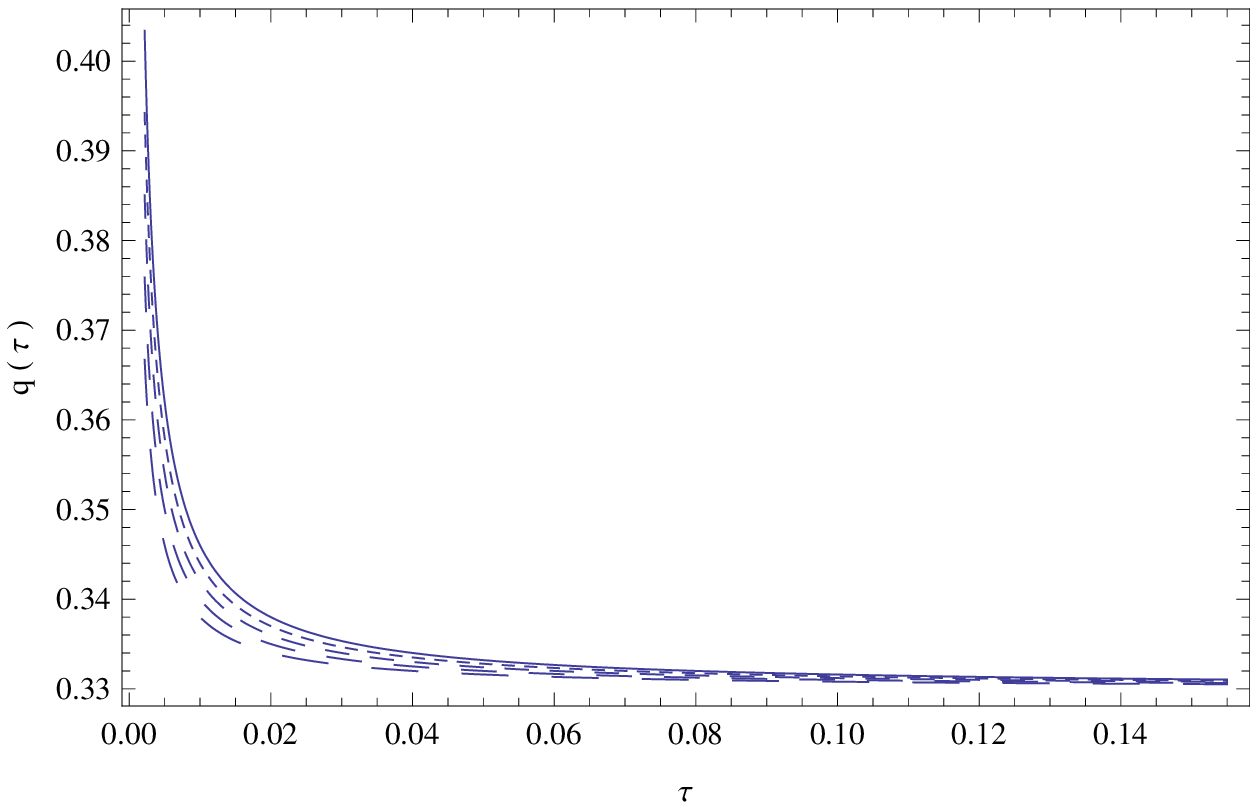}
\includegraphics[width=8.0cm]{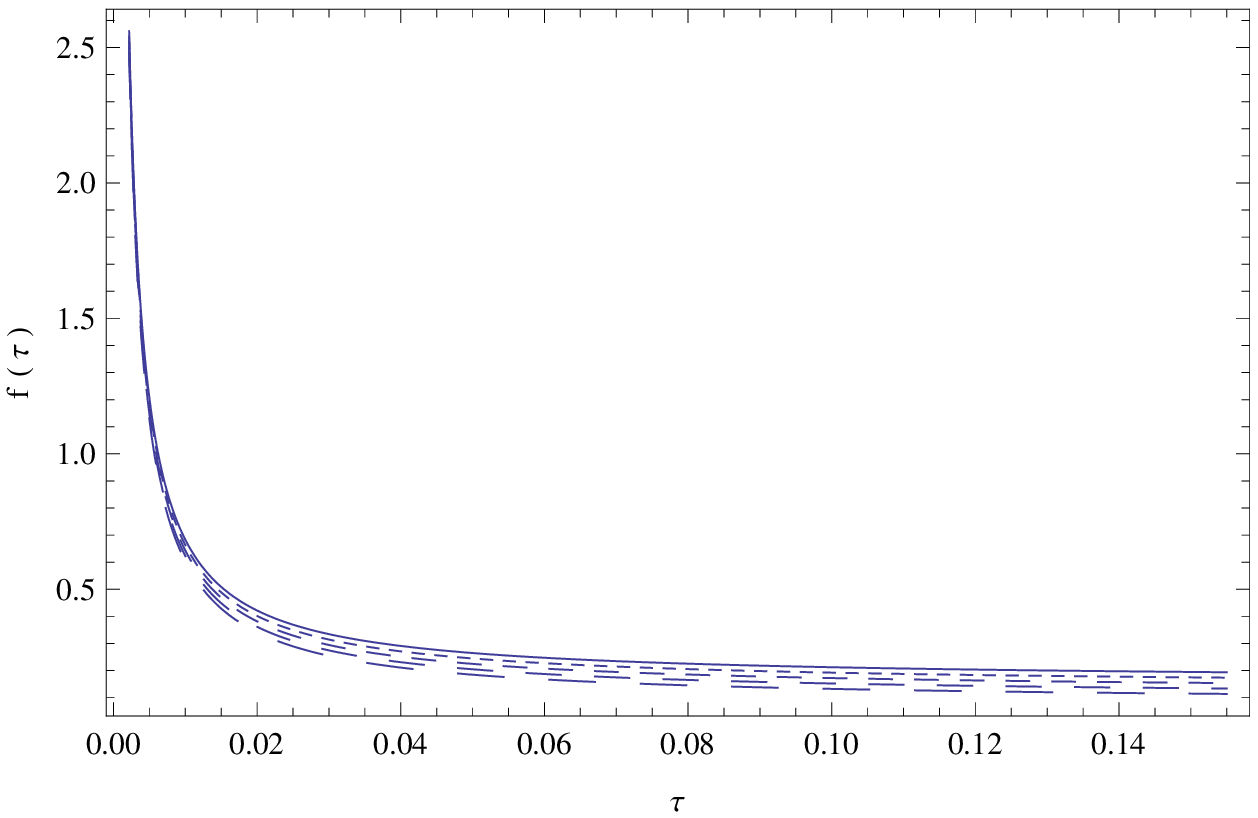}
\caption{Dynamics of the deceleration parameter $q$ (left figure) and of the Bopp-Podolsky vector potential (right figure) of the Bianchi type I dust Universe with Bopp-Podolsky type vector dark energy  for $\tau \geq \tau _{in}=3\times 10^{-5}$, and for different values of the dimensionless Bopp-Podolsky parameter $\mu$: $\mu =160$ (solid curve), $\mu=140$ (dotted curve), $\mu=120 $ (short dashed curve), $\mu =100$ (dashed curve), and $\mu =80$ (long dashed curve), respectively. The initial conditions used to numerically integrate the cosmological evolution equations are $f\left(\tau_{in}\right)=2.4$, $u\left(\tau_{in}\right)=-0.001$, $\sigma \left(\tau_{in}\right)=2.31\times 10^4$, $h\left(\tau _{in}\right)=2.5\times 10^4$, and $r\left(\tau _{in}\right)=1.75\times 10^9$, respectively.}\label{fig6}
\end{center}
\end{figure*}

The cosmological evolution is obtained by numerically integrating Eqs.~(\ref{60})-(\ref{64}) for the zero pressure case, with the  use of the following initial conditions: $f\left(\tau_{in}\right)=2.4$, $u\left(\tau _{in}\right)=-0.001$, $\sigma \left(\tau _{in}\right)=2.31\times 10^4$, $h(\left(\tau _{in}\right)=2.5\times 10^4$, and $r\left(\tau _{in}\right)=1.75\times 10^9$, respectively, and for different values of the Bopp-Podolsky parameter $\mu $. The time variations of the mean Hubble function, matter energy density, shear scalar, anisotropy parameter, deceleration parameter, and of the ratio of the time variation of the Bopp-Podolsky vector potential and scale factor, respectively, are presented in Figs.~\ref{fig4}-\ref{fig6}.

As one can see from Fig.~\ref{fig4}, in the case of the dust Bianchi type I Universe in the presence of a Bopp-Podolsky field, both the mean Hubble function and the matter energy-density are monotonically decreasing functions of time, indicating an expansionary cosmological dynamics. Both the rate of the expansion, as described by the Hubble function, as well as the matter energy density show  almost no $\mu $ in the large time limit. The shear scalar $\sigma$, shown in the left panel of Fig.~\ref{fig5}, is a monotonically decreasing function of $\tau $ during the entire considered period of the cosmological expansion. The time evolution of $\sigma $ is strongly dependent on the numerical values of $\mu $ in the initial and middle stages of evolution. In the long time limit the shear scalar reaches the (approximately) zero value around the time interval $\tau \approx 0.15$, corresponding to an age of the Universe of the order of $t\approx 6.864\times 10^{16}$ s, and to a redshift of $z\approx 2$. The mean anisotropy parameter $A$, represented in the right panel of Fig.~\ref{fig5},  shows a similar behavior, being a monotonically decreasing function for all times. In the long time limit the anisotropy parameter tends to a constant, very small value, which can be approximated as zero for time intervals longer than $\tau =0.15$. For the initial range of considered time intervals  the behavior of the anisotropy parameter depends on the numerical values of $\mu $, but in the long time limit all the models we have considered fully isotropize ($A\approx 0$) independently of the  particular values of $\mu $. 

The deceleration parameter $q$, depicted in the left panel of Fig.~\ref{fig6}, shows that for the considered initial and parameter values the anisotropic dust Universe starts its evolution from a decelerating state with $q\approx 0.4$ at $\tau =3\times 10^{-5}$. Due to the presence of the vector field the deceleration parameter is a monotonically decreasing function of time,  reaching the value $q\approx 0.33$ at $\tau \approx 0.15$.   The overall dynamics of $q$  depends on the numerical values of the Bopp-Podolsky parameter. The evolution of $f$, depicted in the right panel of Fig.~\ref{fig6}, shows a monotonically decrease in time, with the large time dynamics also essentially dependent on the numerical values of $\mu$. In the large time limit $f$ becomes a constant.

Hence an initially anisotropic Bianchi type I Universe fully isotropizes in the presence of a Bopp-Podolsky type vector field, and at $z\approx 2$ the expansion becomes isotropic. However, at the end of the anisotropic era the vector field $f$ still survives as a constant component in the composition of the Universe, and it may play the role of an effective cosmological constant that may trigger the accelerated expansion of the Universe when $z<0.5$.    

\subsection{The effect of the vector field mass on the
cosmological evolution}

In the following we will consider the effect of the mass term $M$ on the
cosmological expansion. In this case, the cosmological equations governing the evolution of the anisotropic Bianchi type I Universe takes the form
\begin{align}\label{67}
\dddot{f}+&(5h+\sigma)\ddot{f}+\f12\sigma(2h-\sigma)\dot{f}+(4h^2+3\dot{h})\dot{f}-\mu^2\dot{f}-(\sigma+2h)(\dot{\sigma}+2\dot{h}+3h\sigma+\mu^2)f-\f12(8h^3+\sigma^3)f-\mu^2 n^2\f{B}{a}=0,
\end{align}
\begin{align}\label{68}
-3h^2+\f34\sigma^2-\f{6\pi}{\mu^2}(\sigma^2+4h^2+4h\sigma)f^2-\f{4\pi}{\mu^2}(\dot{f}-6hf)\dot{f}+\f{4\pi}{\mu^2}(2\ddot{f}+\mu^2f)f+3r+4\pi n^2\f{B^2}{a^2}=0,
\end{align}
\begin{align}\label{69-1}
-2\dot{h}-\dot{\sigma}-3h^2-3h\sigma-&\f34\sigma^2-\f{2\pi}{\mu^2}(\sigma^2+4h\sigma+4h^2)f^2+\f{4\pi}{\mu^2}(\dot{f}+6hf)\dot{f}+\f{4\pi}{\mu^2}(2\ddot{f}+\mu^2f)f-3P-4\pi n^2\f{B^2}{a^2}=0,
\end{align}
and
\begin{align}\label{70-1}
-2\dot{h}+\f12\dot{\sigma}-&3h^2-\f34\sigma^2+\f32h\sigma+\f{4\pi}{\mu^2}(\dot{f}-2f\sigma)\dot{f}+\f{2\pi}{\mu^2}(\sigma^2-8h^2)f^2\nonumber\\&-\f{4\pi}{\mu^2}(2\dot{h}+\dot{\sigma})f^2-\f{4\pi}{\mu^2}(4\dot{f}+f\sigma)hf-4\pi f^2-3P+4\pi n^2\f{B^2}{a^2}=0,
\end{align}
respectively, where a "dot" denotes the derivative with respect to the dimensionless time parameter $\tau$, and we have defined a dimensionless parameter $n$ as $M=nH_0$. Also, we will assume that $\mu\gg1$. Hence the system of Eqs.~(\ref{67})-(\ref{70-1}) can be written as
\be\label{73}
\dot{f}=u,\dot{B}=af,
\ee
\be\label{74}
\dot{a}=\left(h+\frac{2}{\sqrt{3}}\sigma\right)a,
\ee
\begin{align}\label{76}
\dot{u}= \frac{a^2 \left\{8 \pi  f^2 \left[3 (2 h+\sigma
   )^2-2 \mu ^2\right]-96 \pi  f h u+3 \mu ^2 \left(4 h^2-4
   r-\sigma ^2\right)+16 \pi  u^2\right\}-16 \pi  \mu ^2 n^2 B^2}{32 \pi
   a^2 f},
\end{align}
\bea\label{77}
\dot{\sigma}&=&\frac{1}{6 \mu
   ^4 a^2}\Bigg\{a^2 \Bigg[16 \pi  f^2 \left(\mu ^2 ((4 h-\sigma ) (2
   h+\sigma )-3 (P+r))+\mu ^4+8 \pi  u^2\right)+64 \pi ^2 f^4 (2
   h+\sigma )^2+32 \pi  \mu ^2 f u (2 h+\sigma )+\nonumber\\
  && 3 \mu ^4 \left(-6
   h \sigma +4 h^2-4 r-\sigma ^2\right)+16 \pi  \mu ^2
   u^2\Bigg]-16 \pi  \mu ^2 n^2 B^2 \left(8 \pi  f^2+3 \mu ^2\right)\Bigg\},
   \eea
   and
   \bea\label{78}
   \dot{h}&=&\frac{1}{6
   \mu ^4 a(t)^2}\Bigg\{a^2 \Bigg[4 \pi  f^2 \left(\mu ^2 \left(8 h \sigma -4
   h^2+6 (P+r)+5 \sigma ^2\right)-2 \left(\mu ^4+8 \pi
   u^2\right)\right)-32 \pi ^2 f^4 (2 h+\sigma )^2-\nonumber\\
 &&  16 \pi  \mu ^2 f
   u (2 h+\sigma )-3 \mu ^4 \left(2 h^2+3 P+r+\sigma
   ^2\right)+16 \pi  \mu ^2 u^2\Bigg]+64 \pi ^2 \mu ^2 n^2 B^2 f^2\Bigg\},
   \eea
   respectively, while the energy conservation equation becomes
\be\label{79}
   \dot{r}=-3h(r+P).
\ee
   The system of Eqs.~(\ref{73})-(\ref{79}) must be integrated with the initial conditions $f(0)=f_0$, $B(0)=B_0$, $a(0)=a_0$, $u(0)=u_0$, $\sigma (0)=\sigma _0$, $h(0)=h_0$, and $r(0)=r_0$, respectively. Eq.~(\ref{74}) is obtained by eliminating the term $\dot{b}/b$ from the definitions of the Hubble function, $h=(1/3)\left(\dot{a}/a+2\dot{b}/b\right)$, and of the shear scalar, $\sigma ^2=(1/3)\left[\left(\dot{a}/a\right)-\left(\dot{b}/b\right)\right]^2$.

   In the following we will restrict our investigation to the cosmological evolution of the pressureless dust Universe models for a Bopp-Podolsky type dark energy in the presence of a mass term.

 \subsubsection{Cosmological evolution of the anisotropic dust Universe in the presence of a massive Bopp-Podolsky vector field}

In the following we consider the cosmological evolution of a Bianchi type I Universe filled with pressureless dust, with $P=0$, in the presence of a massive Bopp-Podolsky type vector field. In order to facilitate the comparison with the massless case we use the same initial conditions to numerically integrate the cosmological evolution equations (\ref{73})-(\ref{79}), that is,  we adopt as initial conditions $f\left(\tau_{in}\right)=2.4$, $u\left(\tau _{in}\right)=-0.001$, $\sigma \left(\tau _{in}\right)=2.31\times 10^4$, $h(\left(\tau _{in}\right)=2.5\times 10^4$, and $r\left(\tau _{in}\right)=1.75\times 10^9$, respectively, and we consider the same values of $\mu $. For $B$ and $a$ we assume the initial conditions $B\left(\tau _{in}\right)=0.3$ and $a\left(\tau _{in}\right)=0.1$. Moreover, we fix the value of $n$ as $n=2.3$. The time evolutions of the Hubble function, matter energy density, shear scalar, mean anisotropy parameter, deceleration parameter, and Bopp-Podolsky vector potential are presented in Figs.~\ref{fig7}-{\ref{fig9}, respectively.

\begin{figure*}[htp]
\begin{center}
\includegraphics[width=8.0cm]{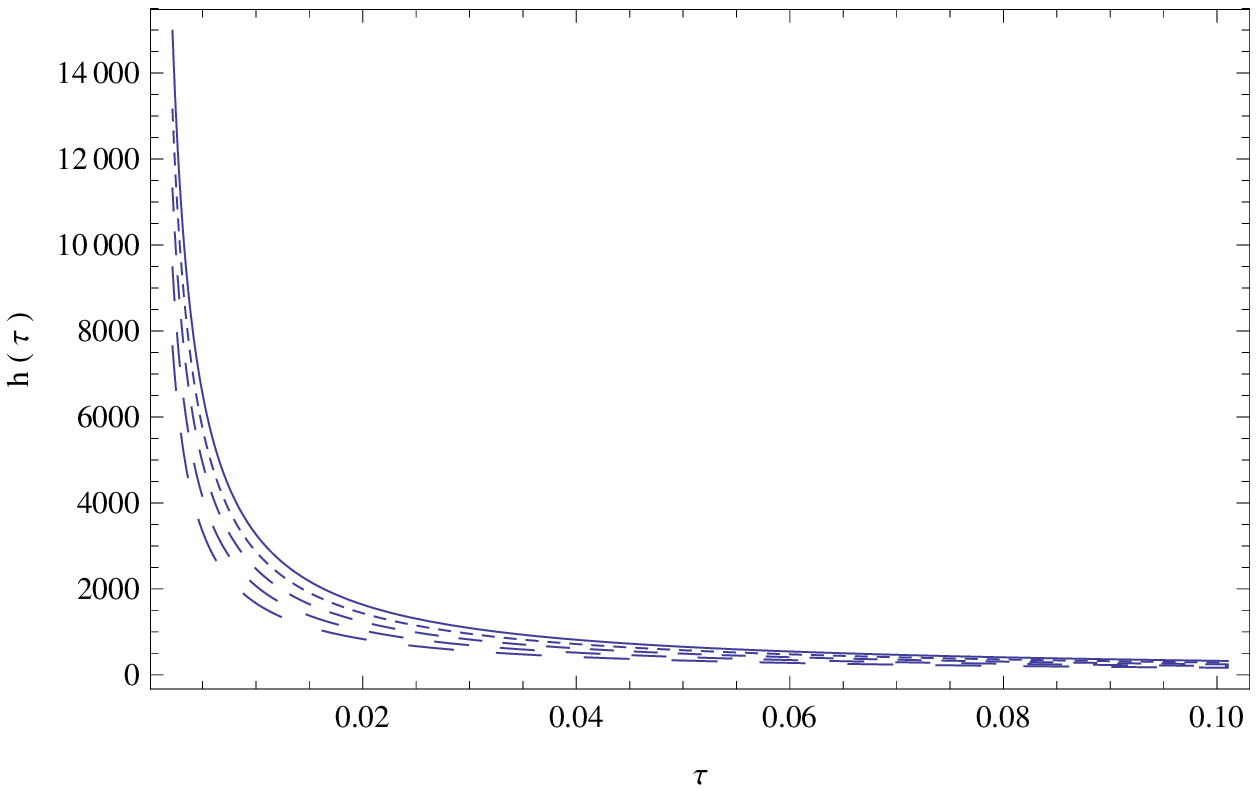}
\includegraphics[width=8.0cm]{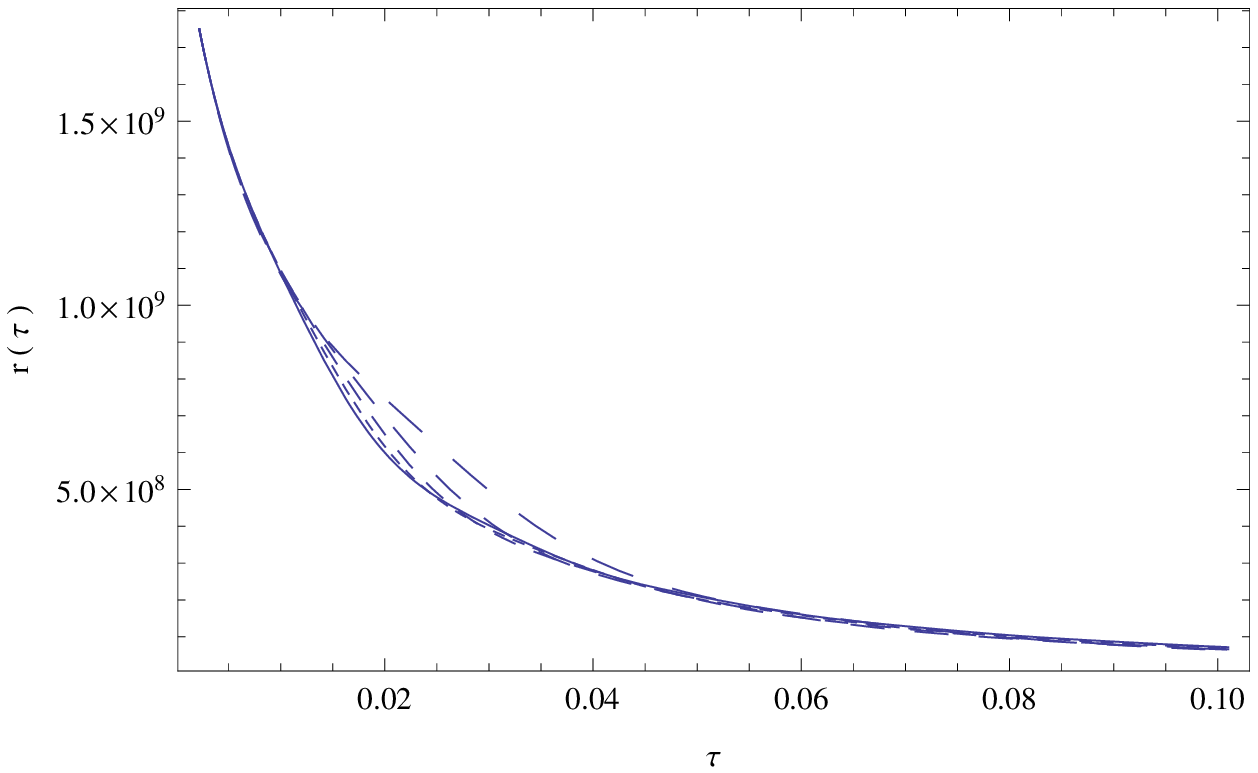}
\caption{Variation of the Hubble function of the anisotropic Bianchi type I dust Universe with
massive Bopp-Podolsky type vector dark energy (left figure), and of the matter energy density (right figure), for $\tau \geq \tau _{in}=3\times 10^{-5}$, and for different values of the dimensionless Bopp-Podolsky parameter $\mu$: $\mu =160$ (solid curve), $\mu=140$ (dotted curve), $\mu=120 $ (short dashed curve), $\mu =100$ (dashed curve), and $\mu =80$ (long dashed curve), respectively. The initial conditions used to numerically integrate the cosmological evolution equations are $f\left(\tau_{in}\right)=2.4$, $u\left(\tau_{in}\right)=-0.001$, $\sigma \left(\tau_{in}\right)=2.31\times 10^4$, $h\left(\tau _{in}\right)=2.5\times 10^4$, and $r\left(\tau _{in}\right)=1.75\times 10^9$,  $B\left(\tau _{in}\right)=0.3$ and $a\left(\tau _{in}\right)=0.1$, respectively. The value of $n$ is fixed as $n=2.3$. }\label{fig7}
\end{center}
\end{figure*}

\begin{figure*}[htp]
\begin{center}
\includegraphics[width=8.0cm]{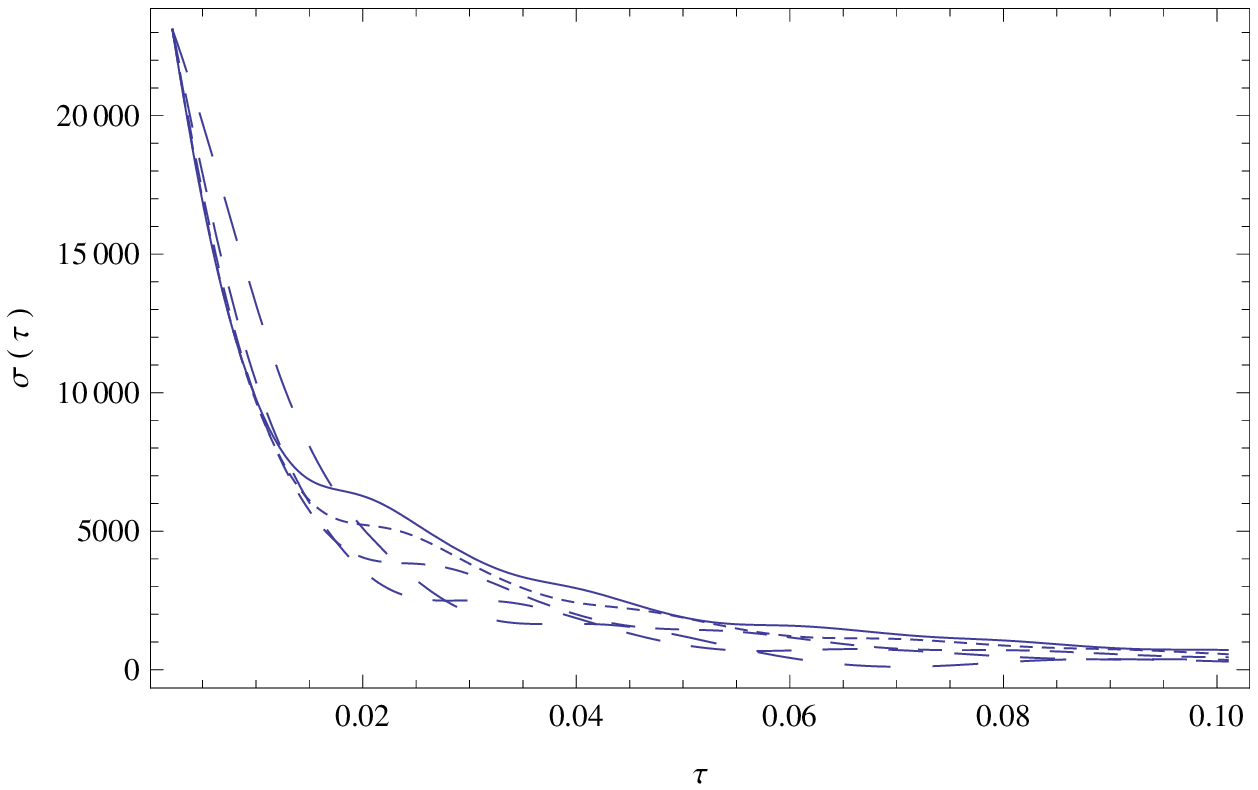}
\includegraphics[width=8.0cm]{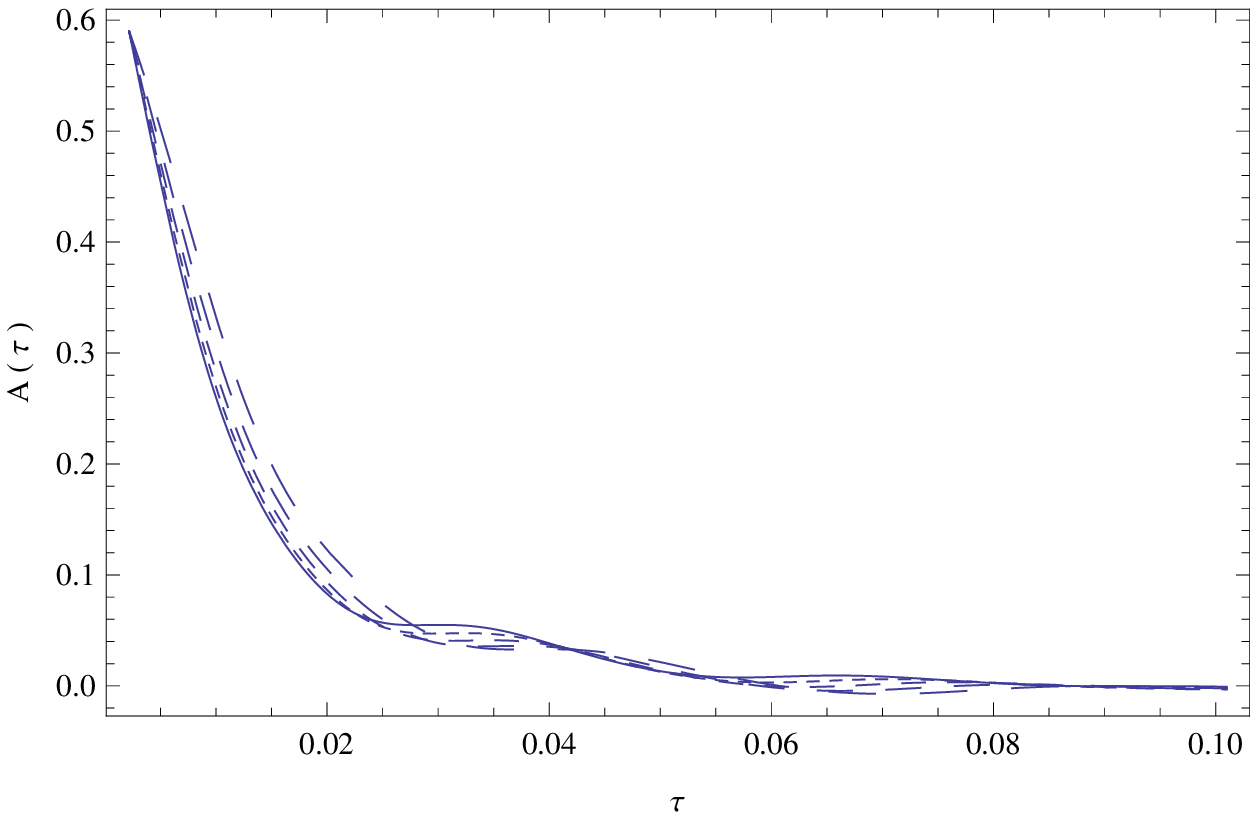}
\caption{Time evolution of the shear scalar $\sigma $ (left figure) and of the anisotropy parameter $A$ (right figure) of the Bianchi type I anisotropic dust Universe with massive Bopp-Podolsky type vector dark energy  for $\tau \geq \tau _{in}=3\times 10^{-5}$, and for different values of the dimensionless Bopp-Podolsky parameter $\mu$: $\mu =160$ (solid curve), $\mu=140$ (dotted curve), $\mu=120 $ (short dashed curve), $\mu =100$ (dashed curve), and $\mu =80$ (long dashed curve), respectively. The initial conditions used to numerically integrate the cosmological evolution equations are $f\left(\tau_{in}\right)=2.4$, $u\left(\tau_{in}\right)=-0.001$, $\sigma \left(\tau_{in}\right)=2.31\times 10^4$, $h\left(\tau _{in}\right)=2.5\times 10^4$, and $r\left(\tau _{in}\right)=1.75\times 10^9$,  $B\left(\tau _{in}\right)=0.3$ and $a\left(\tau _{in}\right)=0.1$, respectively. The value of $n$ is fixed as $n=2.3$. }\label{fig8}
\end{center}
\end{figure*}

\begin{figure*}[htp]
\begin{center}
\includegraphics[width=8.0cm]{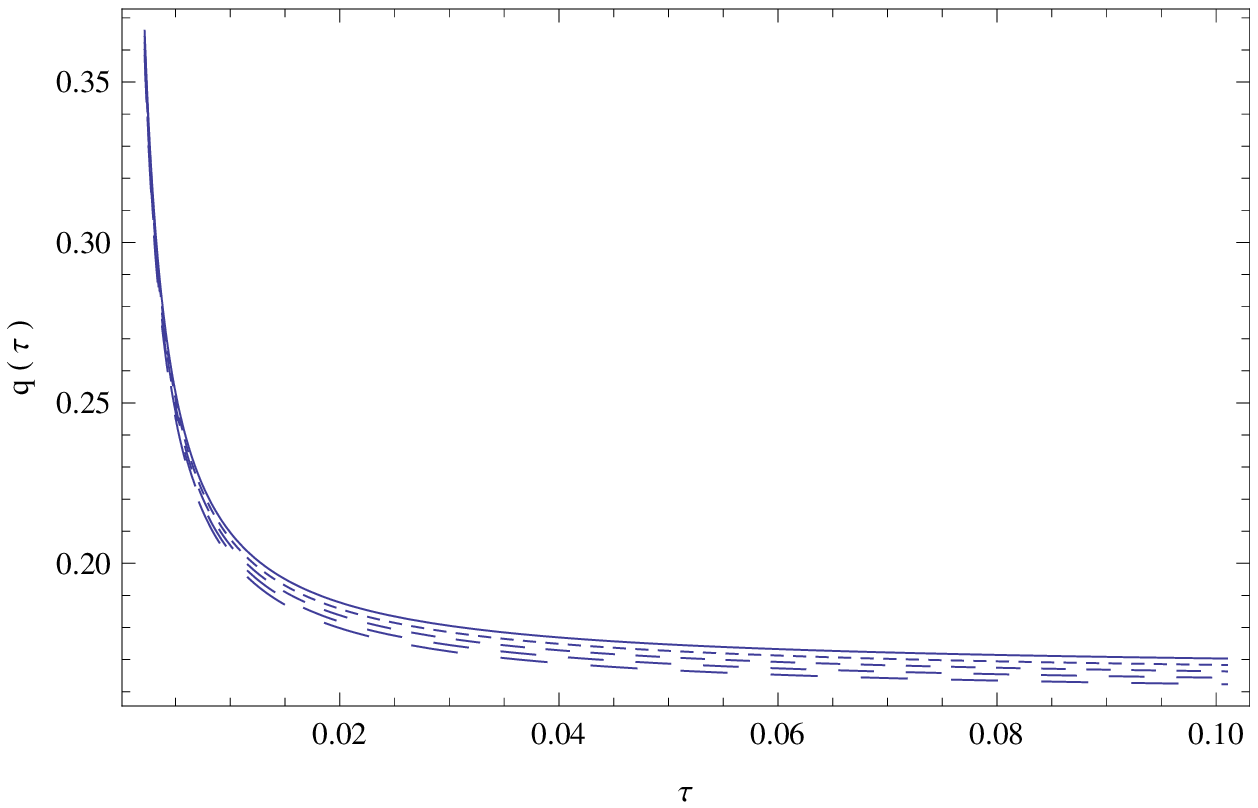}
\includegraphics[width=8.0cm]{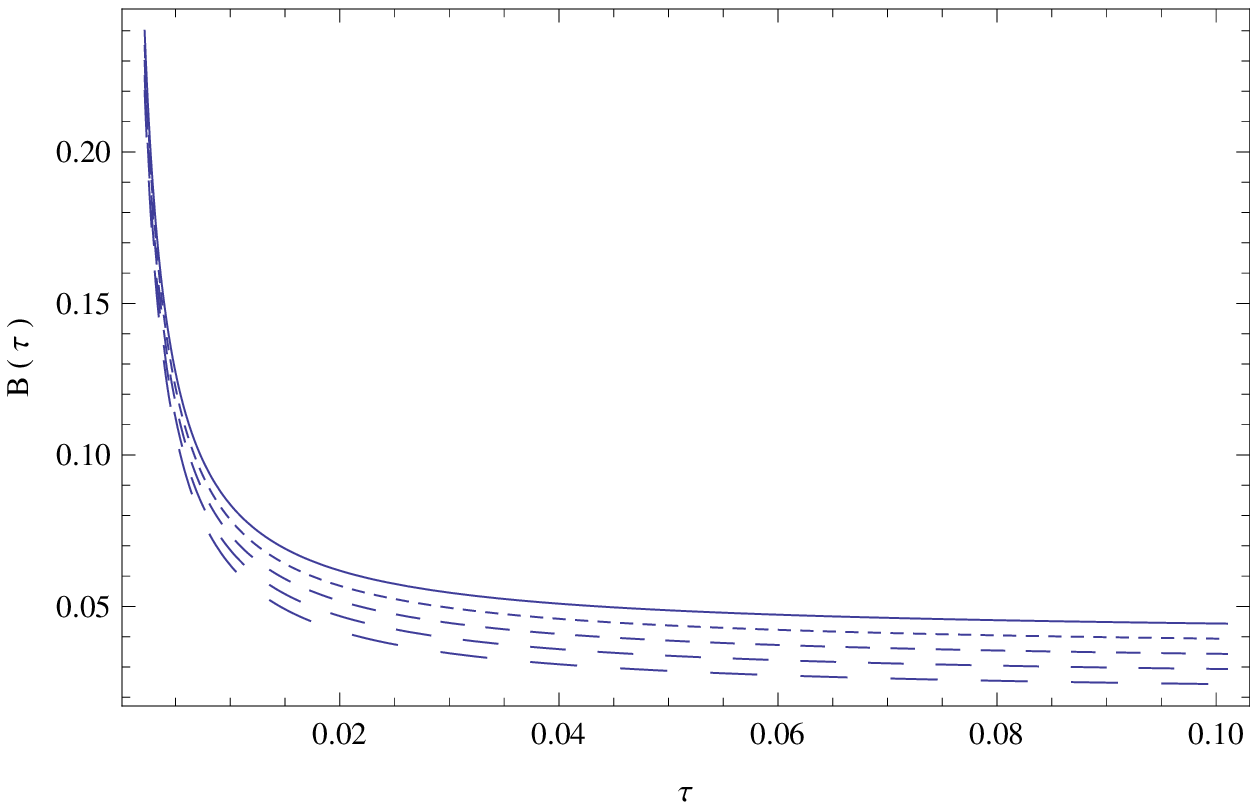}
\caption{Dynamics of the deceleration parameter $q$ (left figure) and of the Bopp-Podolsky vector potential $B$ (right figure) of the Bianchi type I dust  Universe with massive Bopp-Podolsky type vector dark energy  for $\tau \geq \tau _{in}=3\times 10^{-5}$, and for different values of the dimensionless Bopp-Podolsky parameter $\mu$: $\mu =160$ (solid curve), $\mu=140$ (dotted curve), $\mu=120 $ (short dashed curve), $\mu =100$ (dashed curve), and $\mu =80$ (long dashed curve), respectively. The initial conditions used to numerically integrate the cosmological evolution equations are $f\left(\tau_{in}\right)=2.4$, $u\left(\tau_{in}\right)=-0.001$, $\sigma \left(\tau_{in}\right)=2.31\times 10^4$, $h\left(\tau _{in}\right)=2.5\times 10^4$, and $r\left(\tau _{in}\right)=1.75\times 10^9$,  $B\left(\tau _{in}\right)=0.3$ and $a\left(\tau _{in}\right)=0.1$, respectively. The value of $n$ is fixed as $n=2.3$. }\label{fig9}
\end{center}
\end{figure*}

The presence of a mass term for the Bopp-Podolsky vector field has an important influence on the time evolution of the pressureless  Bianchi type I anisotropic cosmological model. Overall, the massive field accelerates the evolution, significantly reducing the time interval necessary to reach the full isotropic phase, thus speeding up  the global isotropization of the Universe. On the other hand, the presence of the mass term does not reduce the dependence of the cosmological evolution on the Bopp=Podolsky parameter $\mu $, at least for the considered range of $\mu $.   The Hubble function, shown in the left panel of Fig.~\ref{fig7}, is a decreasing function of time, with  the rate of expansion of the Universe independent on the numerical values of the model parameter $\mu $ in the large time limit. The energy density of the radiation fluid decreases monotonically during the expansionary phase, and its overall dynamics is not  influenced significantly by the modifications of $\mu $.  The shear scalar $\sigma$, presented in the left panel of Fig.~\ref{fig8},  is a monotonically decreasing, linear function of time, and in the large time limit its behavior is not influenced by the modifications of the numerical values of $\mu $. For the considered range of parameters $\sigma $ becomes (approximately) zero at $\tau \approx 0.10$. The anisotropy parameter $A(\tau)$, shown in the right panel of Fig.~\ref{fig8},  decreases rapidly to zero, indicating that the Universe isotropizes in a shorter time interval, as compared to the massless case. The evolution of $A$ is overall independent on the numerical values of $\mu $. The anisotropic dust Universe accelerates more rapidly as compared to the massless case, and enters the isotropic phase with $q\approx 0.10$, with the deceleration parameter, depicted in the left panel of Fig.~\ref{fig9}, showing a strong dependence on the numerical values of $\mu $.  The Bopp-Podolsky vector field potential $B$, represented in the right panel of Fig.~\ref{fig9}, is a slowly decreasing function of time, also showing a strong dependence on the numerical values of $\mu $.

\section{Discussions and final remarks}\label{sect4}

In this paper we have investigated a specific vector-tensor type gravitational theory which is inspired by the Bopp-Podolsky electrodynamics.
The Bopp-Podolsky theory
represents an interesting generalization of classical electrodynamics which has been extensively investigated at the level of elementary particle theory. Its basic idea, the addition of  quadratic terms in the Maxwell tensor derivatives to the action of the vector-tensor type gravitational models, may prove to be a useful extension of the standard vector-tensor models with minimal coupling between gravity and the vector field. In our model we have also considered the possibility of the existence of some self-interaction processes in dark energy which can be described by means of a potential term $V\left(A^2\right)$. The self interaction potential is a function of the square of the four-potential of the vector type dark energy. In the present analysis we have restricted our investigations to the case of the linear potential only with $V\propto A^2$. A possible non-minimal coupling between the matter current and the four-potential of the vector field was also considered. As compared to the vector models based on standard Maxwell electrodynamics, the addition of the new terms in the action enriches significantly the theoretical framework, thus opening the possibility of a more general approach to vector type dark energy models.

We have considered in detail the cosmological implications of the Bopp-Podolsky type vector dark energy. In our study we have concentrated on the cosmological evolution of the Bianchi type I cosmological models. We have investigated numerically two distinct scenarios, corresponding to the absence and presence of a self-interacting potential for the vector field. In view of the possible applications for the description of the early Universe we have considered first the radiation fluid model. We have investigated it for the case of the massless Bopp-Podolsky vector fields, by varying the numerical values of the parameter $\mu $. The presence of a Bopp-Podolsky type dark energy component could induce a complex dynamical behavior of the early Universe, with the cosmological evolution decelerating ($q>0$). The anisotropy parameter $A$ slowly decreases during the radiation phase. Interestingly, the evolutions of the Hubble function, of the matter energy density and of the shear scalar are not significantly affected by the modifications of the numerical values of the model parameter $\mu $. The evolution is also strongly dependent on the initial conditions adopted for the Hubble function, energy density, shear scalar, and of the function $f$, and its derivatives.

In the case of the dust anisotropic Universe, the cosmological evolution can reach an isotropic phase in both massless and massive cases. The Universe fully isotropizes in both models, so that $A=0$ after a finite time interval, corresponding to $z\approx 2$. The presence of the massive vector field significantly speeds up the cosmological evolution towards an isotropic  phase. The nature of the cosmological evolution strongly depends on the adopted range of the numerical values of the parameter $\mu $ in the massless case, but this dependence is diminished in the presence of the massive vector field. On the other hand, the Bopp-Podolsky vector field  becomes a constant at the end of the anisotropic phase, and its presence could trigger an accelerated, de Sitter type evolution for $z<0.5$.

However, it is important to point out that in the present model the late-time de-Sitter evolution is not an attractor of the system, and that the longtime evolution of the Universe, extending well below the present time, is oscillatory. As shown in Section~\ref{3C}, the period of the cosmological oscillations is of the order of $T\approx (8.88/\mu)t_H$, which for $\mu =10$ is of the same order as the present age of the Universe. Once the Universe reaches this age, the direction of the expansion is reversed, and the Hubble function becomes an increasing function  of the cosmological time, and the Universe  experiences an overall cyclical behavior \cite{osc1,osc2,osc3,osc4,osc5}, consisting of a succession of expanding and collapsing phases. From a theoretical point of view in the present model the classically oscillating solution is obtained by adding a vector type dark energy with quadratic terms in the Maxwell tensor
derivatives to the ordinary baryonic fluid. As a result the  equation of state parameter $w$ satisfies in the long time limit  the condition $w\leq -1$, indicating the possibility of future super-accelerating stages of evolution of the Universe, before its recollapse towards a decelerating phase. One interesting question is if this  type of model can avoid the singularity theorems.  On the other hand we expect that quantum gravity eﬀects, arising from a full theory of quantum gravity, would become important when the size of the Universe approaches the Planck scale. A detailed analysis of the singular behavior of the model would also require the consideration of other quantum effects that may appear due to the presence of quantum fluctuations.  There is also the interesting possibility that after a number of oscillations, the  Universe filled with vector type dark energy with quadratic terms in the Maxwell tensor could evolve to the bounce point through quantum tunneling, and then expand again.  Hence we may tentatively assume that a bouncing Universe
that avoids the classical singularity is also possible in the present model.

Recently,  precise observations of the cosmic microwave background radiation have provided the possibility of testing the fundamental cosmological predictions of inﬂation on the primordial ﬂuctuations,  such as, for example, scale independence and Gaussianity \cite{Planck1}. One of the basic ideas of inflation, also supported by the cosmic no-hair conjecture, conjectures that inﬂation makes classical anisotropy negligibly small. However, recently a number of astrophysical  observations of the large scale structure of the Universe have raised some questions about the validity of the principles of homogeneity and isotropy \cite{An}. Together with the recent Planck results, these observations point towards the possibility of the existence of an intrinsic large scale anisotropy in the Universe. By adopting this line of thought, in \cite{Tomi} it was proposed that the global geometry of the Universe may be described by the homogeneous locally rotationally symmetric (LRS) class of metrics. These interesting geometries induce a preferred direction in the cosmic sky,  and a CMB that is isotropic at the level of the background. Such models can be tested with CMB and supernovae data by using the distortion of the luminosity distances generated by the anisotropic geometry. Hence, by taking into account the latest Planck data, possible existence of small large scale cosmological anisotropies cannot be ruled out completely. However, their physical origins are still unknown, with the most popular explanation being that anisotropy is due to deviations from isotropy of the  primordial fluctuations \cite{Planck1}. But presently no clearly established physical mechanism that could generate such deviations is known \cite{HL}. We suggest that the existence of the Bopp-Podolsky vector dark energy may provide the physical mechanisms that could explain the generation of the anisotropies in the early Universe, as well as their survival at the present time.

Vector field models face the important problem of their stability. Several vector-tensor gravitational theories contain instabilities in the form of
ghosts, or unstable growth of the linearized perturbations \cite{Himm}. The presence of these instabilities is due to the longitudinal
vector polarization modes that appear in the vector-tensor models. The existence of ghosts or tachyons during the early time evolution or in the small wavelength regime of the vector-tensor type cosmological models may indicate that the vacuum of these models is unstable. The problem of stability with respect to small linear perturbations and  the hyperbolicity is also of fundamental importance for the Bopp-Podolsky type dark energy models with derivatives of the Maxwell tensor in the action. A
full solution of this problem can be obtained  by considering  perturbations
of the field equations of our model, and investigating their stability. We will consider this topic in a future publication.

In this paper we have introduced a  vector-tensor type model representing an extension of vector type dark energy models.  Further investigations of the corresponding cosmological models may provide us with some measure  for
discriminating between different evolutionary scenarios suggested by  theoretical structure of the theory. Moreover, this model may contribute to a better understanding of some other fundamental processes like, for example, inflation and
structure formation, which have played a fundamental role in the evolution of our
Universe.

\section*{Acknowledgments}

We would like to thank the anonymous reviewer for comments and suggestions that helped us to significantly improve our manuscript. TH would like to thank  the Yat Sen School of the Sun Yat-Sen University  in Guangzhou, P. R. China, for the kind hospitality offered during the preparation of this work.

\appendix

\section{Variation of the matter current in the interaction term%
}

\label{app}

In this Appendix we present the calculation of  variation of the interaction
term between the matter four-current $j_{\mu}$ and  four-potential of the vector field $A_{\mu}$, which is assumed to be given by
\begin{align}  \label{varap}
	\delta S_{\mathit{int}}&= \delta \int d^4 x \sqrt{-g} A_\mu j^\mu=\int d^4 x
	\delta\sqrt{-g} A_\mu j^\mu+\int d^4 x \sqrt{-g} \left[ \delta A_\mu j^\mu+ A_\mu u^\mu\delta\rho +
	A_\mu \rho \delta u^\mu\right].
\end{align}
Now, we should obtain  variation of the proper density and four velocity
of the perfect fluid. For a perfect fluid we have the constraints
\begin{equation*}
	\delta s=0, \quad \delta n^\mu =0,
\end{equation*}
where $s$ is the entropy density and $n^\mu$ is the baryon number flux
vector density which is defined as
\begin{align}  \label{defn}
	n^\mu=n u^\mu \sqrt{-g},
\end{align}
so that
\begin{align}  \label{numden}
	n=\sqrt{\frac{n^\mu n^\nu g_{\mu\nu}}{g}}.
\end{align}
For the perfect fluid the density is only a function of $n$, i.e. $%
\rho=\rho(n)$. Using thermodynamics relations one  obtains
\begin{equation*}
	\delta\rho=\left(\frac{\partial \rho}{\partial n}\right)_s\delta n=\frac{1}{n%
	}(\rho+p) \delta n.
\end{equation*}
Using equation \eqref{numden} and the above relation one can find
\begin{align}  \label{varr}
	\delta \rho=-\frac{p+\rho}{2}\left(u^\mu u^\nu +g^{\mu\nu}\right)\delta
	g_{\mu\nu}.
\end{align}
To find  variation of the four velocity vector one may use the condition
$\delta n^\mu =0$ and equation \eqref{defn} with the result
\begin{align}  \label{varu}
	\delta u^\mu=\frac{1}{2 } u^\mu u^\alpha u^\beta \delta g_{\alpha\beta}.
\end{align}
Finally, substituting Eqs.~\eqref{varr} and \eqref{varu} into Eq.~%
\eqref{varap} we obtain
\begin{align}
	\delta S_{\mathit{int}}=\int d^4 x \sqrt{-g}\left\{-\frac{1}{2}p A_\mu
	u^\mu(u^\alpha u^\beta +g^{\alpha\beta})+j^\mu \delta A_\mu\right\}.
\end{align}

\begin{thebibliography}{99}
\bibitem{1n} A. G. Riess et al., Astron. J. \textbf{116}, 1009 (1998).

\bibitem{2n} S. Perlmutter et al., Astrophys. J. \textbf{517}, 565 (1999).

\bibitem{3n} R. A. Knop et al., Astrophys. J. \textbf{598}, 102 (2003).

\bibitem{4n} R. Amanullah et al., Astrophys. J. \textbf{716}, 712 (2010).

\bibitem{acc} D. H. Weinberg, M. J. Mortonson, D. J. Eisenstein, C. Hirata,
A. G. Riess, and E. Rozo, Physics Reports \textbf{530}, 87 (2013).

\bibitem{P2} P. A. R. Ade et al., Planck 2013 results. XXVI, arXiv:1303.5086
(astro-ph) (2013); P. A. R. Ade et al., Planck 2013 results. I, arXiv:
1303.5062 [astro-ph) (2013); P. A. R. Ade et al., Planck 2013 results. XVI,
arXiv: 1303.5076 [astro-ph] (2013).

\bibitem{PeRa03} P. J. E. Peebles and B. Ratra, Rev. Mod. Phys. \textbf{75},
559 (2003).
\bibitem{new1} V. Sahni and A. A. Starobinsky, Int. J. Mod. Phys. D \textbf{9}, 373 (2000).
\bibitem{Pa03} T. Padmanabhan, Phys. Repts. \textbf{380}, 235 (2003).

\bibitem{DEreviews} E.~J.~Copeland, M.~Sami and S.~Tsujikawa, Int.\ J.\
Mod.\ Phys.\ D \textbf{15}, 1753 (2006).

\bibitem{Od} S. Nojiri and S. D. Odintsov, Physics Reports \textbf{505}, 59
(2011).

\bibitem{LiM} M. Li, X.-D. Li, S. Wang, and Y. Wang, Frontiers of Physics
\textbf{8}, 828 (2013).

\bibitem{Mort} M. J. Mortonson, D. H. Weinberg, and M. White,
arXiv:1401.0046 (2014).

\bibitem{Amend} L. Amendola, S. Tsujikawa, \textit{Dark Energy, Theory and
Observations}, Cambridge, Cambridge University Press, (2015).

\bibitem{8n} R. Caldwell, R. Dave and P. J. Steinhardt, Phys. Rev. Lett.
\textbf{80}, 1582 (1998).

\bibitem{Fa04} Y. Fujii and K. Maeda, \textit{The Scalar-Tensor Theory of
Gravitation}, Cambridge, Cambridge University Press, (2003); V. Faraoni,
\textit{Cosmology in scalar-tensor gravity}, Dordrecht; Boston, Kluwer
Academic Publishers, (2004).

\bibitem{Tsu} S. Tsujikawa, Class. Quant. Grav. \textbf{30}, 214003 (2013).

\bibitem{11n} L. P. Chimento, A. S. Jakubi and D. Pavon, Phys. Rev. \textbf{%
D 62}, 063508 (2000).

\bibitem{kessence0} T. Chiba, T. Okabe, and M. Yamaguchi,
Phys. Rev. \textbf{D 62}, 023511 (2000); C. Armendariz-Picon, V. F.
Mukhanov, and P. J. Steinhardt,
Phys. Rev. Lett. \textbf{85}, 4438 (2000); C. Armendariz-Picon, V. F.
Mukhanov, and P. J. Steinhardt, 
Phys. Rev. \textbf{D 63}, 103510 (2001); N. Arkani-Hamed, H. C. Cheng, M. A.
Luty, and S. Mukohyama,
JHEP \textbf{0405}, 074 (2004); F. Piazza and S. Tsujikawa,
JCAP \textbf{0407}, 004 (2004).

\bibitem{phan1} R. R. Caldwell, Phys. Lett. \textbf{B. 545}, 23 (2002).

\bibitem{phan2} S. M. Carroll, M. Hoffman, and M. Trodden,
Phys. Rev. \textbf{D 68}, 023509 (2003); P. Singh, M. Sami, and N. Dadhich,
Phys. Rev. \textbf{D 68}, 023522 (2003); M. Sami and A. Toporensky,
Mod. Phys. Lett. \textbf{A 19}, 1509 (2004); J. M. Cline, S. Jeon, and G. D.
Moore, 
Phys. Rev. \textbf{D 70}, 043543 (2004); E.~Elizalde, S.~Nojiri and
S.~D.~Odintsov, Phys. Rev. \textbf{D 70}, 043539 (2004); E.~Elizalde,
S.~Nojiri, S.~D.~Odintsov, D.~Saez-Gomez and V.~Faraoni, Phys.\ Rev. \textbf{%
D 77}, 106005 (2008).

\bibitem{phan3} A. Yu. Kamenshchik, Class. Quantum Grav. \textbf{30}, 173001
(2013).

\bibitem{phan4} U. Alam, V. Sahni, T. D. Saini, and A. A. Starobinsky, Mon.
Not. Roy. Astron. Soc. \textbf{354}, 275 (2004).
\bibitem{new2} B. Boisseau, G. Esposito-Farese, D. Polarski and A. A. Starobinsky, Phys. Rev. Lett. 85, 2236 (2000).
\bibitem{Bu70} W. Hu and I. Sawicki, Phys. Rev. D 76, 064004 (2007); S. A. Appleby and R. A. Battye, Phys. Lett. B 654, 7 (2007);  A. A. Starobinsky, JETP Lett. 86, 157 (2007); H. A. Buchdahl, Mon. Not. Roy. Astron. Soc. \textbf{150}, 1
(1970); A. De Felice and S. Tsujikawa. Living Rev. Rel. \textbf{13}, 3
(2010); T. P. Sotiriou and V. Faraoni, Rev. Mod. Phys. \textbf{82}, 451
(2010); S. Nojiri and S. D. Odintsov, Phys. Rept. \textbf{505}, 59 (2011);
F. S. N. Lobo, arXiv:0807.1640 [gr-qc]; S. Capozziello and M. De Laurentis,
Phys. Rept. \textbf{509}, 167 (2011); G.~J.~Olmo, Int.\ J.\ Mod.\ Phys.\
\textbf{D 20}, 413 (2011).

\bibitem{mime} S. Nojiri, S. D. Odintsov, Mod. Phys. Lett. A 29, 1450211 (2014); G. Leon, E. N. Saridakis, JCAP 1504, 031 (2015); Z. Haghani, S. Shahidi, M. Shiravand, arXiv:1507.07726 [gr-qc]; R. Myrzakulov, L. Sebastiani, S. Vagnozzi, Eur. Phys. J. C 75, 9 (2015).

\bibitem{Har1} O. Bertolami, C. G. Boehmer, T. Harko, and F. S.N. Lobo,
Phys. Rev. \textbf{D 75}, 104016 (2007); T. Harko, Phys. Lett. \textbf{B 669}%
, 376 (2008); T. Harko and F. S. N. Lobo, Eur. Phys. J. \textbf{C 70}, 373
(2010); T. Harko, F. S. N. Lobo, and O. Minazzoli, Phys. Rev. \textbf{D 87},
047501 (2013).

\bibitem{Har2} T. Harko, F. S.N. Lobo, S. Nojiri, and S. D. Odintsov, Phys.
Rev.\textbf{D 84}, 024020 (2011); T. Harko, Phys. Rev. {\bf D 90}, 044067 (2014).

\bibitem{WCW} Z. Haghani, T. Harko, H. R. Sepangi, and S. Shahidi, JCAP
\textbf{10}, 061 (2012); Z. Haghani, T. Harko, H. R. Sepangi, and S. Shahidi, Phys. Rev. D 88, 044024 (2013).

\bibitem{Har3} T. Harko, T. S. Koivisto, F. S. N. Lobo, and G. J. Olmo,
Phys. Rev. \textbf{D 85}, 084016 (2012); N. Tamanini and C. G. B\"{o}hmer,
Phys. Rev. \textbf{D 87}, 084031 (2013).

\bibitem{Har4} Z. Haghani, T. Harko, F. S. N. Lobo, H. R. Sepangi, and S.
Shahidi, Phys. Rev. \textbf{D 88}, 044023 (2013).

\bibitem{EIBI} A. S. Eddington, \textit{The Mathematical Theory of Relativity%
}, Cambridge University Press, Cambridge, UK, 1924; M. Born and L. Infeld,
Proc. R. Soc. Lond. \textbf{A 144}, 425 (1934); S. Deser and G. W. Gibbons,
Class. Quant. Grav. \textbf{15}, L35 (1998); M. Banados and P. G. Ferreira,
Phys. Rev. Lett. \textbf{105}, 011101 (2010).

\bibitem{HT} T. Harko, F. S. N. Lobo, G. Otalora, and E. N. Saridakis, JCAP
\textbf{12}, 021 (2014).
\bibitem{vgb} A. Oliveros, Enzo L. Solis, Mario A. Acero, Mod. Phys. Lett. A 31, 1650009 (2015).
\bibitem{Revn} T. Harko and F. S. N. Lobo, Galaxies \textbf{2}, 410 (2014).

\bibitem{Revn1} S. Capozziello, T. Harko, T. S. Koivisto, F. S. N. Lobo, and
G. J. Olmo, Universe \textbf{1}, 199 (2015).

\bibitem{v1} C. Armendariz-Picon, JCAP \textbf{0407}, 007 (2004).
\bibitem{new5} R. C. G. Landim, Eur. Phys. J. C 76, 480 (2016).

\bibitem{v2} V. V. Kiselev, Class. Quantum Grav. \textbf{21}, 3323 (2004);
H. Wei and R.-G. Cai, Phys. Rev. \textbf{D 73}, 083002 (2006); T. S.
Koivisto and D. F. Mota, J. Cosmol. Astropartic. Phys. \textbf{0808}, 021
(2008); J. Beltr\'{a}n Jim\'{e}nez and A. L. Maroto, Phys. Rev. \textbf{D 78}%
, 063005 (2008); J. Beltr\'{a}n Jim\'{e}nez, R. Lazkoz, and A. L. Maroto,
Phys. Rev. \textbf{D 80}, 023004 (2009); V. V. Lasukov, Russian Physics
Journal \textbf{53} 296 (2010); E. Carlesi, A. Knebe, G. Yepes, S.
Gottloeber, J. Beltr\'{a}n Jim\'{e}nez, and A. L. Maroto, Monthly Not. Royal
Astron. Soc. \textbf{418}, 2715 (2011); E. Carlesi, A. Knebe, G. Yepes, S.
Gottloeber, J. Beltr\'{a}n Jim\'{e}nez, Antonio L. Maroto, Monthly Not.
Royal Astron. Soc. \textbf{424}, 699 (2012); N. Br\'{e}ton, Accelerated
Cosmic Expansion, Astrophysics and Space Science Proceedings, Volume 38,
Springer International Publishing Switzerland, p. 61, (2014).

\bibitem{v3} C. G. B\"ohmer and T. Harko, Eur. Phys. J. \textbf{C 50}, 423
(2007).

\bibitem{SupracondDE} S-D. Liang and T. Harko,
Phys. Rev. \textbf{D 91}, 085042 (2015).

\bibitem{supobs} Z. Keresztes, L. A. Gergely, T. Harko, and S.-D. Liang,
Phys. Rev. \textbf{D 92}, 123503 (2015).

\bibitem{Bopp} F. Bopp, Ann. Phys. (Leipzig) \textbf{38}, 345 (1940).

\bibitem{Podolsky} B. Podolsky, Phys. Rev. \textbf{62}, 68 (1942).

\bibitem{applBopp} S. I. Kruglov, J. Phys. \textbf{A 43}, 245403 (2010); R.
B. B. Santos, Modern Physics Letters \textbf{A 26}, 1909 (2011); P. Gaete,
International Journal of Modern Physics \textbf{A 27}, 1250061 (2012); A. E.
Zayats, Ann. Phys. \textbf{342}, 11 (2014); V. Perlick, arXiv:1411.0286
(2014); J. Gratus, V. Perlick, and R. W. Tucker, J. Phys. A: Math. Theor.
\textbf{48}, 435401 (2015).

\bibitem{LaLi} L. D. Landau and E. M. Lifshitz, The classical theory of fields, Oxford, Pergamon Press, United Kingdom 1971
\bibitem{def} C.-M. Chen, T. Harko, and M. K. Mak, Phys. Rev. {\bf D 62},  124016 (2000); T. Harko and M. K. Mak, Class. Quant. Grav. {\bf 21},  1489 (2004).

\bibitem{Planck1} P. A. R. Ade et al., Planck 2015 results. XX. Constraints on inflation, Astronomy \& Astrophysics {\bf 594}, A20 (2016).

\bibitem{An}  C. J. Copi, D. Huterer, D. J. Schwarz, and G. D. Starkman, Adv. Astron. {\bf 2010}, 847541 (2010).

\bibitem{Mukh} V.  Mukhanov, Physical foundations of Cosmology, Cambridge University Press,  Cambridge, 2005

\bibitem{osc1} E. R. Harrison, Monthly Not. Royal Astron. Soc. {\bf 137},  69 (1967).

\bibitem{osc2}  M. P. Dabrowski,  Annals Phys. {\bf 248}, 199 (1996).

\bibitem{osc3} M. Novello and S. E. Perez Bergliaﬀa,  Physics Reports {\bf 463}, 127 (2008).

\bibitem{osc4}  P. W. Graham, B. Horn, S. Kachru, S. Rajendran, and Gonzalo Torroba,  JHEP {\bf 1402}, 029 (2014).

\bibitem{osc5} J. Kehayias and  R. J. Scherrer, JCAP {\bf 12} 015   (2015).

\bibitem{Tomi}  T. S. Koivisto, D. F. Mota, M. Quartin, and T. G. Zlosnik, Phys. Rev. {\bf D 83}, 023509 (2011).

\bibitem{HL} T. Harko and F. S. N. Lobo, JCAP {\bf 07},  036 (2013).

\bibitem{Himm}  B. Himmetoglu, C. R. Contaldi, and M. Peloso, Phys.
Rev. Lett. {\bf 102}, 111301, (2009); B. Himmetoglu, C. R.
Contaldi, and M. Peloso, Phys. Rev. {\bf D 80}, 123530 (2009).

\end{thebibliography}
\end{document}